\documentclass{aa}  
\usepackage{graphicx}
\usepackage{txfonts}
\usepackage{natbib}
\usepackage{amssymb}
\newcommand{\mk}{\rm Mrk~71}
\newcommand{\hi}{H {\small I}}
\newcommand{\hii}{H~\texttt{II}~}

\newcommand{\hb}{H$\beta$}

\newcommand{\ha}{H$\alpha$}

\begin{document}

   \title{An IFU investigation of possible Lyman continuum escape from Mrk 71/NGC 2366\thanks{Based on observations collected at the Centro Astron{\'{o}}mico Hispanico en Andaluc{\'{\i}}a (CAHA) at Calar Alto, operated jointly by the Andalusian Universities and the Instituto de Astrof{\'{\i}}sica de Andaluc{\'{\i}}a (CSIC)}}
\titlerunning{IFU investigation of Mrk 71}

   \author{Genoveva Micheva\inst{1},
     Edmund Christian Herenz\inst{2},
     Martin M. Roth\inst{1},
     G\"oran \"Ostlin\inst{2},
          \and  Philipp Girichidis\inst{1}
          }

   \institute{Leibniz-Institut f\"ur Astrophysik, An der Sternwarte 16, D-14482 Potsdam, Germany\\
              \email{gmicheva@aip.de}
         \and
             Department of Astronomy, Oskar Klein Centre for Cosmoparticle
             Physics, Stockholm University, AlbaNova University Centre, 106 91 Stockholm, Sweden\\
             }
   \authorrunning{Micheva et al.}
   \date{Received December 12, 2018; accepted February 11, 2019}

 
  \abstract
   {Mrk 71/NGC 2366 is the closest Green Pea (GP) analog and candidate Lyman
     Continuum (LyC) emitter. Recently, $11$ LyC-leaking GPs have been
     detected through direct observations of the ionizing continuum, making
     this the most abundant class of confirmed LyC-emitters at any redshift. High
     resolution, multi-wavelength studies of GPs can lead to an
     understanding of the method(s), through which LyC escapes from these galaxies.}
   {The proximity of Mrk 71/NCG 2366 offers unprecedented detail on the
     inner workings of a GP analog, and enables us to identify the mechanisms of LyC escape.}
   {We use $5825\mbox{-}7650$\AA\ integral field unit PMAS observations to study the
     kinematics and physical conditions in Mrk 71. An 
     electron density map is obtained from the [S {\small II}] ratio. A fortuitous
     second order contamination by the [O {\small II}] $\lambda3727$ doublet
     enables the construction of an electron temperature map. Resolved maps
     of sound speed, thermal broadening, ``true'' velocity dispersion, and Mach
     number are obtained and compared to the high resolution
     magneto-hydrodynamic SILCC simulations.}
   {Two regions of increased velocity dispersion indicative of
     outflows are detected to the north and
     south of the super star cluster, knot B, with redshifted and blueshifted
     velocities, respectively. We confirm the presence of a faint broad kinematical
     component, which is seemingly decoupled from the outflow regions, and is
     fainter and narrower than previously reported in the literature. Within
     uncertainties, the low- and high-ionization gas move together. Outside of
     the core of Mrk 71, an increase in Mach numbers is detected, implying a
     decrease in gas density. Simulations suggest this drop in density can be
     as high as $\sim4$ dex, down to almost optically thin levels, which
     would imply a non-zero LyC escape fraction along the outflows even when
     assuming all of the detected H {\small I} gas is located in front of Mrk
     71 in the line of sight.}  
   {Our results strongly indicate that kinematical feedback is an important
     ingredient for LyC leakage in GPs. }


   \maketitle
%

\section{Introduction}
Lyman continuum (LyC) radiation, escaping from star-forming galaxies (SFGs) into
the intergalactic medium, is likely responsible for the reionization of the
universe at redshifts $z\gtrsim6$
\citep[e.g.,][]{Razoumov2010,Alvarez2012,Bouwens2015,Duncan2015,Bouwens2016}. At such
epochs bright active galactic nuclei (AGN) may be too few to contribute
significantly \citep[e.g,][]{Meiksin2005,Fontanot2012,Fontanot2014}, and the
average ionizing emissivity of the more numerous faint AGN is observed to be
insufficient \citep[e.g.,][]{Micheva2017a,Parsa2018}. 

Low redshift ($z\lesssim3$) LyC emitters have proven difficult to find
\citep[e.g.,][]{Leitherer1995,Bergvall2006,Siana2007,Siana2015,Vanzella2010,Leitet2011,Leitet2013} due to the
attenuation of LyC photons by the intergalactic medium (IGM). For decades,
the only confirmed, directly detected LyC escape from a low-redshift galaxy
was that in Haro 11 \citep{Bergvall2006, Leitet2011}, with \citet{Ostlin2015}
providing the first spatially resolved kinematic study of a confirmed LyC emitter. Only recently
has significant progress been made in detecting local LyC emitters. Green Peas
are the most prolific class of local LyC-leaking galaxies, showing a very high 
LyC detection rate. So far, $11$ out of 11 GPs have been confirmed as leaking LyC by direct
detection with HST/COS \citep{Izotov2016a,Izotov2016b,Izotov2018a,Izotov2018b}. GPs are compact galaxies which appear green in SDSS $g$, $r$, $i$ composite images due to their extremely high equivalent width of [O \small{III}] $\lambda\lambda4959$, $5007$ and their redshifts of only $z\sim0.2\mbox{-}0.4$. They have sub-solar metallicities, low masses, 
and high specific star-formation rate, and represent good analogs to high
redshift SFGs \citep[e.g.,][]{Cardamone2009,Izotov2011,Bian2016}. While much closer than their high-z cousins,
GPs are nevertheless at distances that still present severe difficulties for
a spatially resolved analysis of their properties. Pioneering work with integral
field unit (IFU) observations of four GPs \citep{Lofthouse2017} revealed that
two are rotationally supported, with seemingly undisturbed morphology, and two
are dispersion-dominated, leading the authors to conclude that mergers may not
be a necessary driver of their star formation properties, and by extension,
for their LyC emission. At the same time, \citet{Amorin2012} find complex,
multi-component H$\alpha$ line profiles, with implied high-velocity
gas in all five of their observed GPs. While these GPs are not confirmed LyC emitters, and therefore their kinematical properties cannot be taken as a necessary condition for LyC escape, outflows nevertheless seem to play a role in enabling LyC leakage. For example, \citet{Chisholm2017} find that outflows can be detected in all LyC emitters they examine, albeit with velocities not statistically different from a control sample of non-leakers. Further, \citet{Martin2015} find that ULIRG galaxies with the strongest outflows manifest the lowest column densities of neutral gas. Simulations also support the scenario of outflows enabling LyC escape. \citet{Trebitsch2017} find that mechanical feedback from supernovae explosions has a vital role in disrupting dense star-forming regions and in clearing low-density escape paths for the LyC photons. While IFU observations of a large sample of GPs would help settle some of these issues, the large distances to these galaxies would likely prohibit the detailed identification and characterization of the exact
escape mechanisms of LyC. Much closer galaxies, showing the same characteristics as GPs, are required for such an endeavor.

Mrk 71/NGC 2366 is one such galaxy. It is the closest GP analog and a LyC
candidate, at a distance of $3.4$ Mpc, i.e., close enough for the detection of
individual stars. The NGC 2366 galaxy hosts the giant H {\small II} region Mrk
71, which dominates the ionization properties of the galaxy, and drives the
similarities to GPs \citep{Micheva2017b}. The integrated properties of Mrk 71/NGC 2366 are fully
consistent with GPs in terms of morphology, excitation properties, specific
star-formation rate, kinematics, absorption of low-ionization species, dust
reddening, and chemical abundances. One notable difference is that Mrk 71/NGC
2366 is $\sim2$ orders of magnitude fainter than GPs, and hence could
represent the faint end of the GP luminosity distribution. This galaxy has been studied in terms
of gas and stellar kinematics
\citep[e.g.,][]{Roy1991,Roy1992,GonzalezDelgado1994,Hunter2001,Hunter2012,Binette2009},
reddening and abundances \citep[e.g.,][]{Izotov1997,Hunter1999,James2016}, UV spectral properties
\citep{Rosa1984,Leitherer2011}, continuum and emission line fluxes
\citep[e.g.,][]{Hunter1999,Moustakas2006,Kennicutt2008,Dale2009}, H {\small I} properties
\citep{Hunter2012}, resolved stellar populations
\citep[e.g.,][]{Thuan2005,McQuinn2010}, and ionization  
properties \citep[e.g.,][]{Izotov1997,Drissen2000,James2016,Sokal2016}. 

\begin{figure*}
     \begin{center}
       \includegraphics[height=8cm]{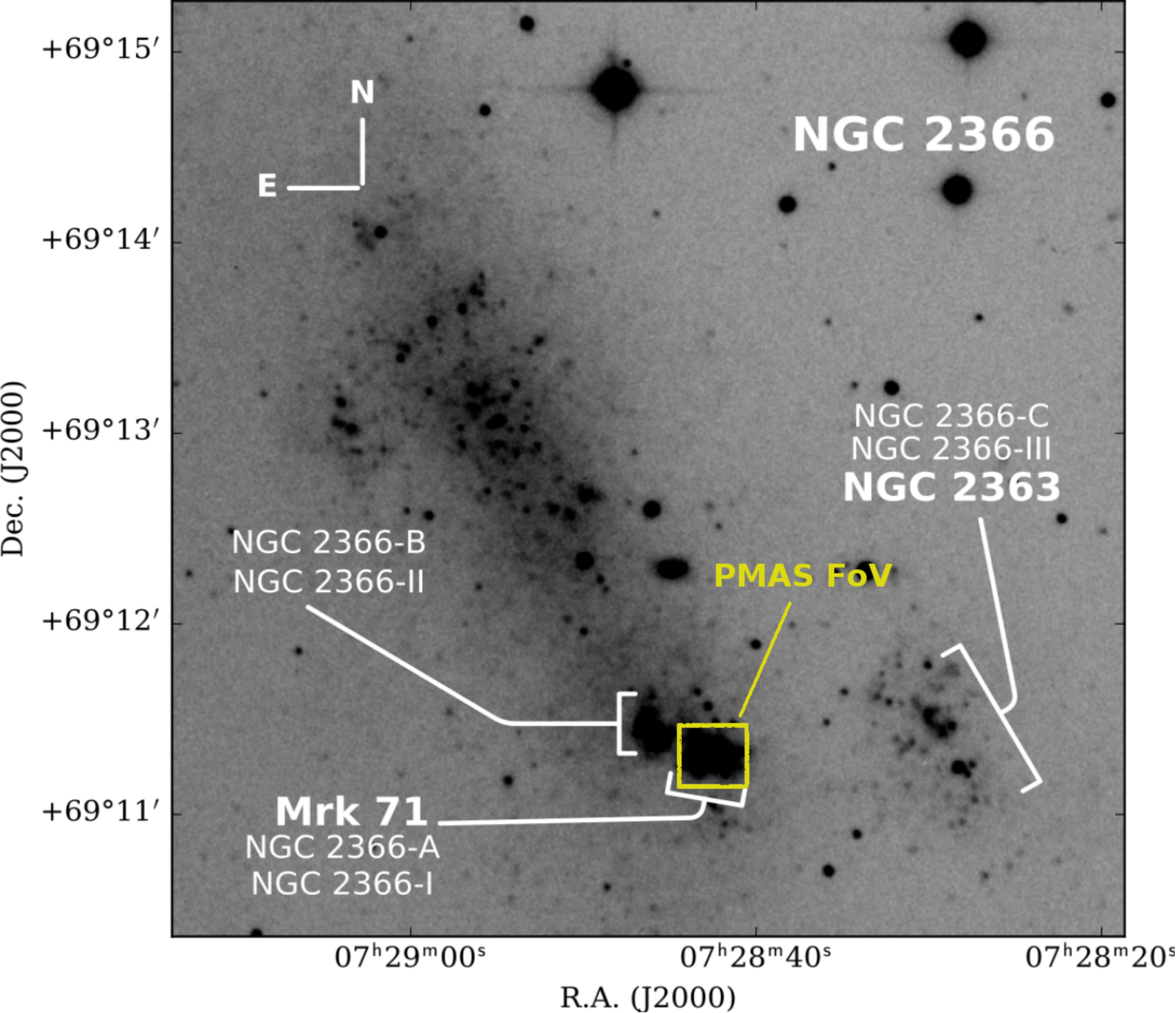}
       \includegraphics[width=\hsize]{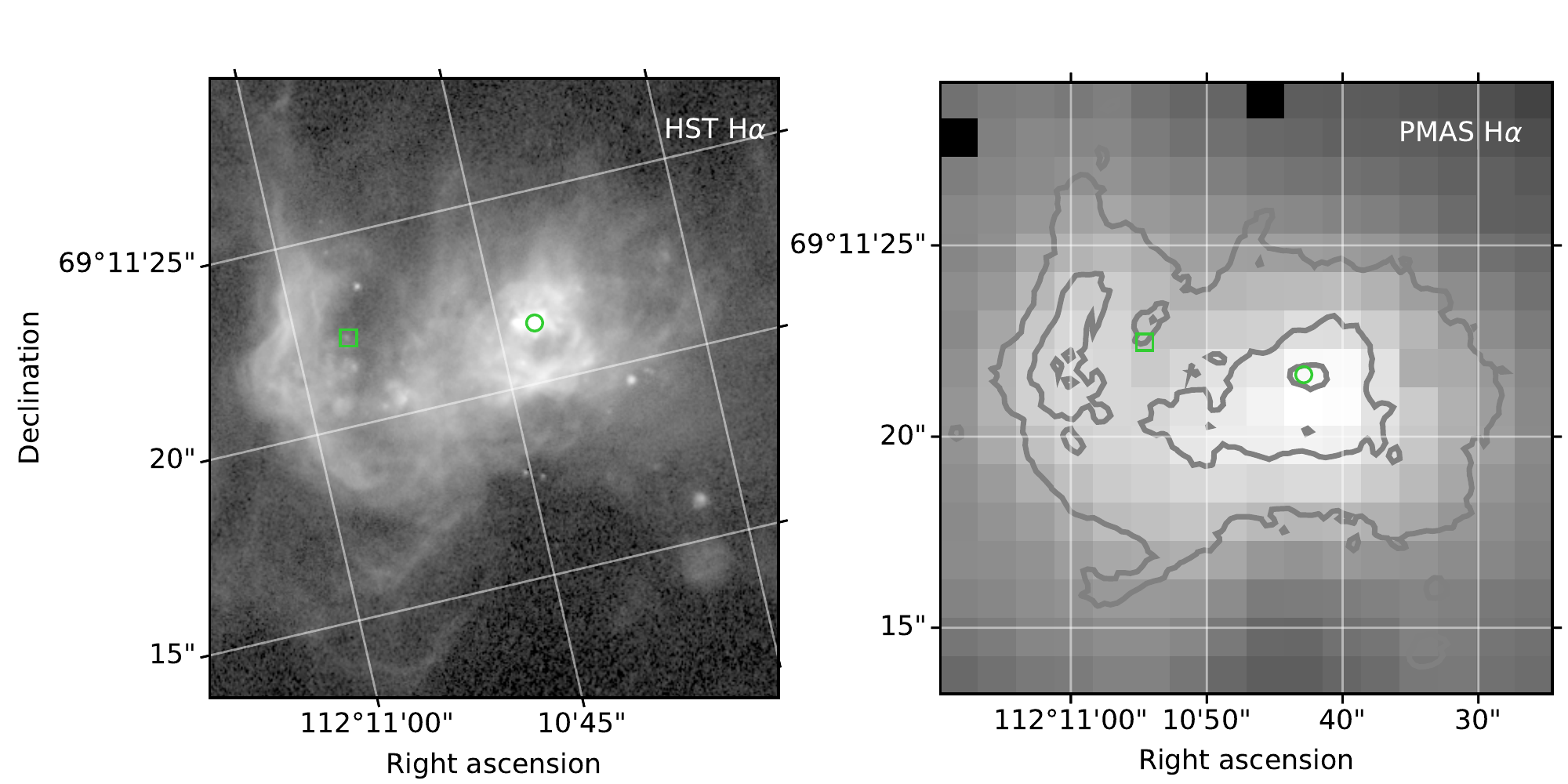}
       \caption{Top row: NGC 2366, with labeled major substructures and the PMAS field of view. Figure
         reproduced from \citet{Micheva2017b}. 
         Bottom row: Continuum-subtracted \ha\ images from the HST (left) and PMAS
         (right). The positions of knots A and B are indicated with open
         circle and square, respectively. HST \ha\ iso-intensity contours at $8.1\times10^{-6}$, $1.3\times10^{-6}$ and $1.9\times10^{-7}$ erg/s/cm$^2$/\AA/arcsec$^2$, are overplotted on the PMAS
         image for orientation. The position angle of the HST field is rotated
         by $\sim14\deg$ relative to the PMAS image.} 
       \protect\label{fig:Ha_images} 
     \end{center}
   \end{figure*}

In this work, we present the first IFU data of the Mrk 71 region, taken with the fiber-fed Potsdam
Multi-Aperture Spectrophotometer \citep[PMAS;][]{Roth2005} with the field
of view coverage shown in Figure \ref{fig:Ha_images} in H$\alpha$. For
orientation, also shown is HST H$\alpha$ imaging from \citet{James2016} of
approximately the same region, and the location of the two super star clusters
is marked - knots A and B in the nomenclature of \citet{Micheva2017b}. We use
the integral field unit data of PMAS to perform a spatially resolved study of
the kinematical properties and physical conditions in Mrk 71, with a focus on
characterizing the channels, through which LyC photons may be escaping. 

%

The layout of the paper is as follows. In Section \ref{sec:data} we present
the PMAS observations, reductions, and final data product. The H$\alpha$
kinematics are analyzed in Section \ref{sec:kinematics}, and we take a closer
look at the spatially resolved physical conditions in Section
\ref{sec:physcond}. A discussion of the observational results and comparison
to simulations in presented in Section \ref{sec:discussion}, and the
conclusions summarized in Section \ref{sec:conclusions}. The Appendices
\ref{sec:UVflat} and \ref{sec:dustmap} contain details about the reductions
and the continuum subtraction of auxiliary narrowband imaging,
respectively. Throughout this paper, the AB magnitude system is used.  

%

\section{IFU observations of Mrk 71}\protect\label{sec:data}
We obtained $2\times1800$ seconds of on-target exposure with the fiber-fed Potsdam Multi-Aperture
Spectrophotometer \citep[PMAS;][]{Roth2005} at the Calar Alto $3.5$m telescope
during the night of March 25th $2017$ (PMAS run362). We used a setup similar
to the observations presented in \citet{Herenz2016}, namely, the R1200
backward-blazed grating with the Lens Array (LArr), with a field of view of
$16\arcsec\times16\arcsec$ in double magnification mode. Arc and continuum lamp calibration and
bias exposures were taken at the same position in the sky, before and after the Mrk 71 exposures, as well as a
separate sky exposure of $300$ seconds, taken between the two Mrk 71
exposures, and offset by $1\deg$ in right ascension and declination from the
Mrk 71 coordinates. The spectro-photometric standard star G191B2B \citep{Oke1990} was observed for the
purpose of flux calibration, with arc lamp, continuum lamp, and bias exposures
taken at its position in the sky. For the duration of the observations, the
sky conditions were clear and stable. The observation summary is shown in Table \ref{tab:obslog}.

%
%
\begin{table*}
\caption{Observation summary for the night of 2017-03-25. }             
\label{tab:obslog}      
\centering          
\begin{tabular}{cclll}     
\hline\hline       
ID & $t_{\rm exp.}$ & Seeing         & Airmass & Comment\\
   &      [s]     &  FWHM$[\arcsec]$ & &\\
\hline                    
Mrk 71 & $2\times1800$& $1.7^{\arcsec}$ &  $1.19$ & \hii\ region\\
sky  & $300$          & $1.7^{\arcsec}$ &  $1.19$& $\Delta \alpha,\Delta \delta=(+1^{\arcmin},-1^{\arcmin})$    \\
G191B2B&$300$        & $1.6^{\arcsec}$ &  $1.12$& standard star\\
\hline                  
\end{tabular}
\tablefoot{Field of view: $16\arcsec\times16\arcsec$; Wavelength range:
  $5825\mbox{-}7650$ \AA; Resolving power at $6500$ \AA: $\bar{R}_{\rm
    6500\AA}=\lambda/\Delta\lambda =4704$ }
\end{table*}

\subsection{Data reduction and calibration}\protect\label{sec:reduction}
The data were reduced following the recipe outlined in \citet{Sandin2010} and \citet{Herenz2016},
using the \texttt{p3d}-package\footnote{\url{http://p3d.sourceforge.io}}
\citep{Sandin2010,Sandin2012}. The reduction steps are given in detail by
these authors, and we only summarize them briefly below. A master bias was
created from average stacking of five bias images with the
\texttt{p3d}-subroutine \texttt{p3d\_cmbias}. Care was taken to select only
those frames which had the same median bias level as that measured in the
overscan regions of each science frame and each chip. The \texttt{p3d}
sub-routine \texttt{p3d\_ctrace} was then used to obtain a trace mask for each
science exposure from the associated continuum lamp exposures. Similarly,
individual dispersion masks were obtained for each science exposure from the
associated arc lamp observations, using the sub-routine
\texttt{p3d\_cdmask}. The resulting wavelength sampling was $0.46\AA$ per pixel.
We used the modified optimal extraction (MOX) setting with the subroutine
\texttt{p3d\_cobjex} to extract the spectra. Cross-talk contamination between  
neighboring spectra for the LArr IFU is negligible, and we have not applied a
correction, as recommended by \citet{Sandin2010}. A correction for
fiber-to-fiber throughput variations was applied by flatfielding the
data. Individual flatfields for each science and arc lamp frames were created from the
associated continuum lamp observations, using the \texttt{p3d\_cflatf}
sub-routine.

The same reduction and extraction procedures were applied to the standard star
data. A sensitivity function was then obtained from the latter, using the
sub-routine \texttt{p3d\_fluxsens}. Here we assumed the standard airmass
extinction curve at Calar-Alto \citep[table 5 in ][]{Sanchez2007}. The science
exposures of \mk\ were then separately flux calibrated using this sensitivity
function. The final data product is a data cube of $16\times16\times3991$
volume pixels (voxels), with a spatial sampling of $1\arcsec\times 1\arcsec$
spatial pixels (spaxels), and with spectral sampling of $0.46$ \AA\ per pixel,
covering the wavelength range from  $5825$ to $7650$ \AA. The wavelength
calibration has an absolute accuracy of $0.1$, $0.05$, and $0.07$\AA\ at
$\lambda=5850$, $6500$, and $7440$\AA, respectively, which corresponds to an
accuracy of $\pm 5$, $\pm 2.5$, and $\pm 2.8$ km/s. These uncertainties were
obtained by measuring the wavelength at the peak position of HgNe arc lines at
the indicated wavelengths, and taking the average offset between these measurements
and their theoretical values. Vignetting at the
blue and red ends extends to $\pm350$ \AA\ \citep{Roth2010}, and makes
those parts of the array unusable. This affects the He {\small I}
$\lambda5875.6$ and [O {\small II}] lines, for which the first and last $16$ fibers along the spatial direction in the row-stacked spectra\footnote{These are 2D arrays in which each row corresponds to an individual one-dimensional spectrum.} are affected by vignetting and not used in the analysis. These fibers are mapped to the first and last columns of the 2D spatial array.

We examined the effect of differential atmospheric refraction (DAR) with the
subroutine \texttt{p3d\_darc}. At the bluest and reddest wavelengths of our
spectra, the DAR correction amplitude reaches at most $0.1''$, which is
negligible compared to our resolution of $1''$ per spaxel. We therefore
omitted applying a DAR correction in order to avoid another interpolation of
the data.

\subsection{Calibration of the second order [O {\small II}] $\lambda3727$ emission}\protect\label{sec:UVOII}
Our data were taken with the PMAS setup intended
for observations of the extended Lyman $\alpha$ Reference Sample (eLARS)
(Herenz et al., in prep.; Bridge et al., in prep). The redshifts of
the eLARS galaxies are such that a UV-blocking filter was not inserted, because
the second order contamination by UV lines like [O {\small II}] $\lambda3727$
is redshifted outside of the spectral window. Mrk 71 is, however, at a
distance of only $3.4$ Mpc, and therefore a second order spectrum of the UV [O
  {\small II}] doublet is indeed present at the expected position of
$2\times\lambda=7454\AA$ and with twice the doublet wavelength separation.

The second order contamination of the UV [O {\small II}] is treated as a
fortuitous feature throughout this work. We flux calibrate the UV [O {\small II}] doublet
with HST imaging in F373N from the archive (PID 13041, PI: B. James). The HST
filter contains both doublet lines, and therefore we can only calibrate the
sum of [O {\small} II] $\lambda3726+3729$. The standard reduction steps
described in Section \ref{sec:reduction} do not account for the fact that the
second order UV spectrum needs to be flatfield corrected with a UV
flatfield. With the second order UV arc lamp line of Hg$\lambda3650.14$\AA, we
create the UV sensitivity map shown in Figure \ref{fig:UV_hg} in the Appendix, and use it to
flatfield the UV [O {\small II}]. 

\subsection{Sky subtraction}

The sky subtraction method that gave the lowest residuals at line center was
directly subtracting the available sky exposure, although due to the shorter exposure
time of the latter, this increased the noise in the continuum of the
sky-subtracted spectra. We only aim to reliably trace the gas emission lines,
and therefore the increased noise in the continuum and in the outskirts of the
field of view is of no consequence to our analysis. In addition, we tested
other methods for removing the sky emission. We tried modeling the sky 
lines with Gaussian functions and the sky continuum with a low order polynomial and
subtracting the composite model from each spectrum, and also modeling and subtracting
only the sky continuum, and accounting for sky lines by fitting them simultaneously with
the gas emission lines. These methods performed worse, and we opted for
directly subtracting the flux-calibrated sky exposure from the two science
exposures as the best method of removing sky emission. The science exposures
were average-stacked after the sky subtraction. 

\subsection{Measurement of gas emission lines}
\begin{figure*}
     \begin{center}
       \includegraphics[width=\hsize]{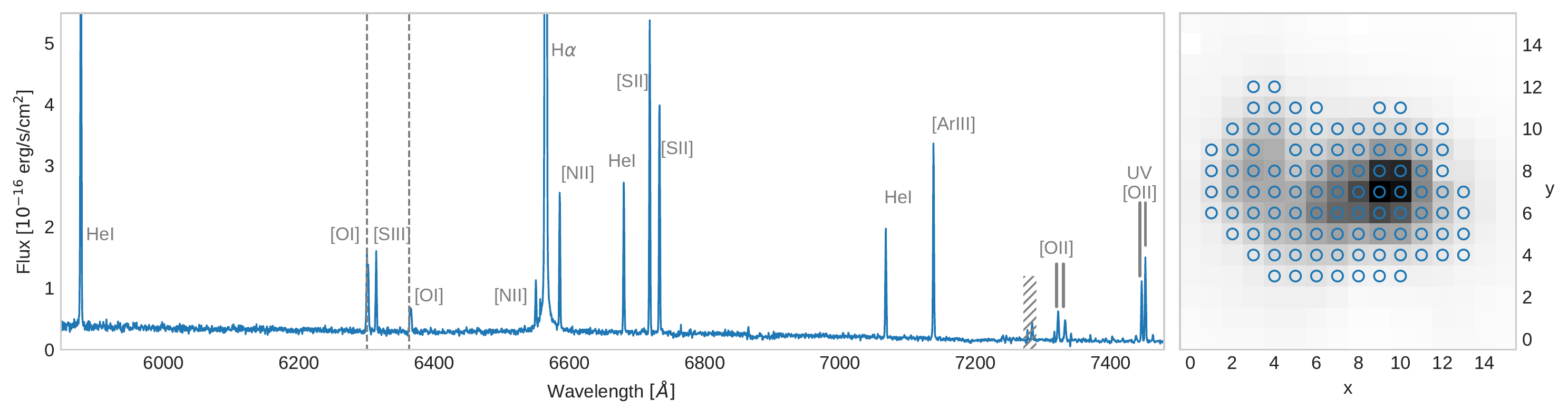}
       \caption{Sky-subtracted and flux-calibrated average spectrum from the
         PMAS data cube (left panel) of the region
         marked in the \ha\ intensity map (blue circles in right panel,
         at the center of each spaxel). All identified lines are labeled,
         including the UV [O {\small II}] second order spectrum. The sky
         emission at [O {\small I}] $\lambda\lambda6300,6363$ (gray dashed
         lines) could not be fully subtracted with the sky exposure. The
         y-axis is truncated in order to better show weak lines like [O
           {\small II}] $\lambda\lambda7320,7330$. Vignetting affects the
         regions $\lambda\lesssim6175$ \AA\ and $\lambda\gtrsim7300$ \AA\ at
         the edges of the detector (columns $x=0$ and $x=15$).}
       \protect\label{fig:fullspec} 
     \end{center}
   \end{figure*}

\protect\label{sec:linemeasures}
We identified the following lines in the first order spectra of \mk: He {\small I}
$\lambda5875.6,\ \lambda6678.1,\ \lambda7065.19$, O {\small I}
$\lambda\lambda 6300,6363$, [S {\small III}] $\lambda6312$, [N {\small II}]
$\lambda\lambda 6548,6584$, \ha, [S {\small II}] $\lambda\lambda 6716,6731$,
[Ar {\small III}] $\lambda 7135.8$, [O {\small II}]
$\lambda\lambda7320,7330$. 
Figure \ref{fig:fullspec} shows the full spectral range and labeled lines for our
target. The second order spectrum is labeled as UV [O {\small II}]. Columns
with index $\mathrm{x=0}$ and $\mathrm{x=15}$ are affected by vignetting in
the wavelength range $\lambda\lesssim6175$ \AA\ and $\lambda\gtrsim7300$ \AA\,
respectively. Throughout the analysis, these columns are discarded for the
emission lines that fall within these wavelength ranges.

%

The flux, width, and central wavelength of each line, along with the
uncertainties in these quantities, were obtained via Monte-Carlo (MC)
simulations in the following way. The \texttt{p3d} subroutine
\texttt{p3d\_ifsfit} measures the flux, full width at half maximum (FWHM), and
the wavelength at line peak by fitting a Gaussian function to the line, where the position,
amplitude and width of the lines are allowed to vary during the fit. This
first estimate provides fitting errors for each quantity. In each MC run, we
perturbed each spectrum by adding random noise to each line spectrum, drawn
from the (assumed normal) distribution of the original \texttt{p3d\_ifsfit}
error. Each perturbed spectrum was passed to \texttt{p3d\_ifsfit} to obtain
the flux, width, and central wavelengths for that MC run. For a given
spectrum, the average of $1000$ MC realizations are taken as our best
estimates of the line flux, FWHM, and wavelength. For each of these
properties, the standard deviation of the distribution of $1000$ MC runs is
taken as the associated $1\sigma$ uncertainty.

Despite our sky subtraction, residuals of the oxygen sky emission lines at
$6300$ and $6363\AA$ remained. Due to the low redshift of \mk, these lines are
only $\sim1.5\AA$ away from the [O {\small I}] nebular emission lines of our
target. For these two cases, we therefore fitted the residuals of the sky
lines simultaneously with the target lines. 

\subsection{Resolving power}
\begin{figure}
     \begin{center}
       \includegraphics[width=\hsize]{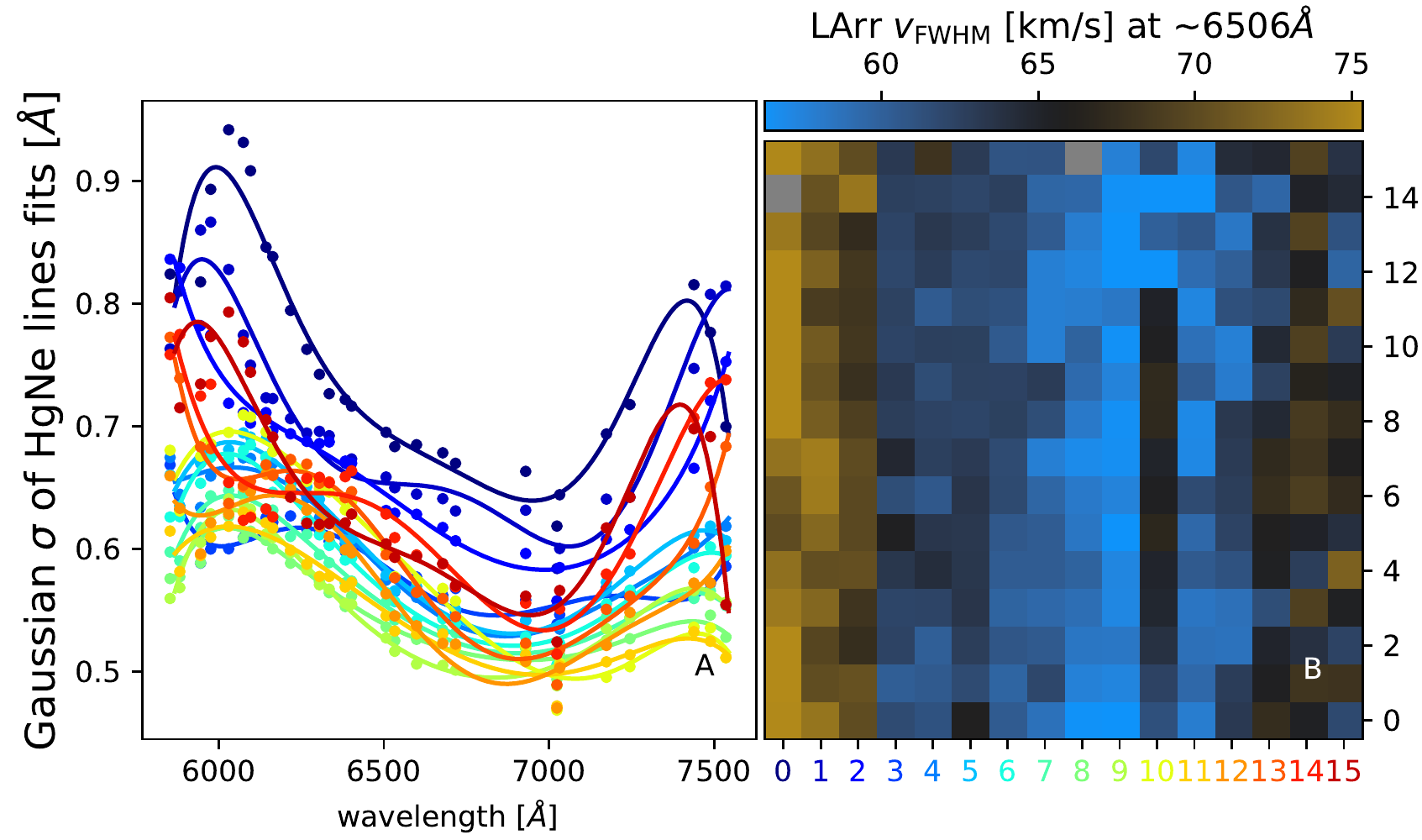}
       \caption{ (A) Instrumental broadening for $26$ HgNe arc lines (filled circles), here for representation purposes obtained by stacking the spectra along columns in (B). The data points and the ID of the column that produced them have the same color-coding. 
         The colored solid lines are a sixth order polynomial fit, used to interpolate the
         resolving power at the desired wavelength. (B):
         An example resolving power map close to \ha. The x-axis
         shows the column ID, color-coded as in (A). }\protect\label{fig:respow}
     \end{center}
   \end{figure}

To obtain the instrumental line broadening for every line and every spaxel,
we used the subroutine \texttt{p3d\_ifsfit} on the wavelength-calibrated arc
lamp exposures to measure the widths of all arc lamp lines in every spaxel. We
have defined the resolving power as $R=\lambda/\Delta\lambda$, where
$\Delta\lambda=\mathrm{FWHM}$. A total of $26$ HgNe lines in the range
$5852\mbox{-}7536$ \AA\ were fitted with Gaussians. In order to obtain
resolving power maps at the wavelengths of the gas emission lines of our
target, we fitted a sixth order polynomial to the arc line widths to enable
interpolation. This is illustrated in Figure  \ref{fig:respow}a, where for
plotting purposes we have stacked all spectra in each column in order to avoid
cluttering. A resolving power map at the wavelength of \ha\ is shown in Figure
\ref{fig:respow}b. At this wavelength the median $v_{FWHM}$ is $63$ km/s,
corresponding to a resolving power of $\bar{R}_{FWHM}$ of $4704$. 

%

\subsection{Internal reddening correction}\protect\label{sec:redcor}
In order to correct our spectra for internal dust extinction in Mrk 71 we use
archival HST data. This is necessary, since our spectral range does not cover
the H$\beta$ line. To map the dust in Mrk 71 we use available HST imaging in
F656N (H$\alpha$), F487N (H$\beta$), F814W (H$\alpha$ continuum) and F438W
(H$\beta$ continuum) from the archive (PID 13041, PI: B. James). A continuum
scaling factor for each filter, constant across the field of view, was
obtained with the mode method \citep{Keenan2017}, using the residual mode
representation of \citet{Micheva2018} to identify the scaling factor. Further
details on the mode method are provided in Section \ref{sec:dustmap} of the Appendix. Ideally,
one would want to determine a spatially resolved scaling factor, in order to
account for possible color gradients due to different stellar
populations. This can be done by modeling the spectral energy distribution
(SED) at each spaxel and from the model determining the continuum at line
position \citep[e.g.,][]{Hayes2009}. We choose to use the mode method because
it is simpler, the dust reddening is known to be overall low in Mrk 71
\citep[e.g.,][]{Izotov1997}, and none of our results rely crucially on a
highly accurate reddening correction. 
%

Once obtained, the continuum-subtracted H$\alpha$/H$\beta$ map is then
convolved with the average seeing of our observations ($1.7''$, Table \ref{tab:obslog}), and resampled
to the resolution of PMAS of $1''$ per spaxel, using the python package
\texttt{reproject} \citep{pythonReproject}. Mode residual plots, indicating the scaling factors for
H$\alpha$ and H$\beta$, are shown in Figure \ref{fig:mu}, and the resulting reddening
map is shown in Figure \ref{fig:dustmap} in the Appendix. Continuum-subtracted H$\alpha$
contours are overplotted, tracing the $8.1\times10^{-6}$, $1.3\times10^{-6}$ and $1.9\times10^{-7}$ erg/s/cm$^2$/\AA/arcsec$^2$ iso-intensity levels.
The location of the two super star clusters, knots A and B, discussed at length in
\citet{Micheva2017b}, are indicated for orientation. 

The H$\alpha$/H$\beta$ map is used to correct the spectrum in each spaxel for
internal reddening. We assume the \citet[][R$\mathrm{_V}=3.1$]{Cardelli1989}
reddening law and use the python software \texttt{pyneb}
\citep{Luridiana2015} to obtain the reddening correction at each wavelength
for case B recombination (T$_e=10^4$ K, n$_e=100$ cm$^{-3}$). We will measure
T$_e$ and n$_e$ explicitly in Section \ref{sec:temden}, however, the dependence
of \ha/\hb\ on the temperature and density is weak, and the dust map we obtain
should not be significantly affected by departures from the assumed T$_e$ and
n$_e$ values. For now we note that, assuming the same reddening law, the
resulting extinction correction c(\hb) is consistent with the literature,
suggesting that our continuum subtraction is acceptable. For example, with the
\citet{Fitzpatrick1999} reddening law, at knot A we obtain c(\hb)$=0.31$, while
\citet{James2016} reported $0.2\mbox{-}0.35$, using a spatially
resolved continuum scaling factor. A consistent value was obtained from \citet{GonzalezDelgado1994}, using long-slit spectroscopy. For the region around knot B our dust map shows no dust, with \ha/\hb\ ratios below $2.86$. This is consistent with a long-slit spectroscopy measurement by \citet{Izotov1997}, who find c(\hb)=0.01 for knot B.

%
   \begin{figure*}
     \begin{center}
       \includegraphics[width=\hsize]{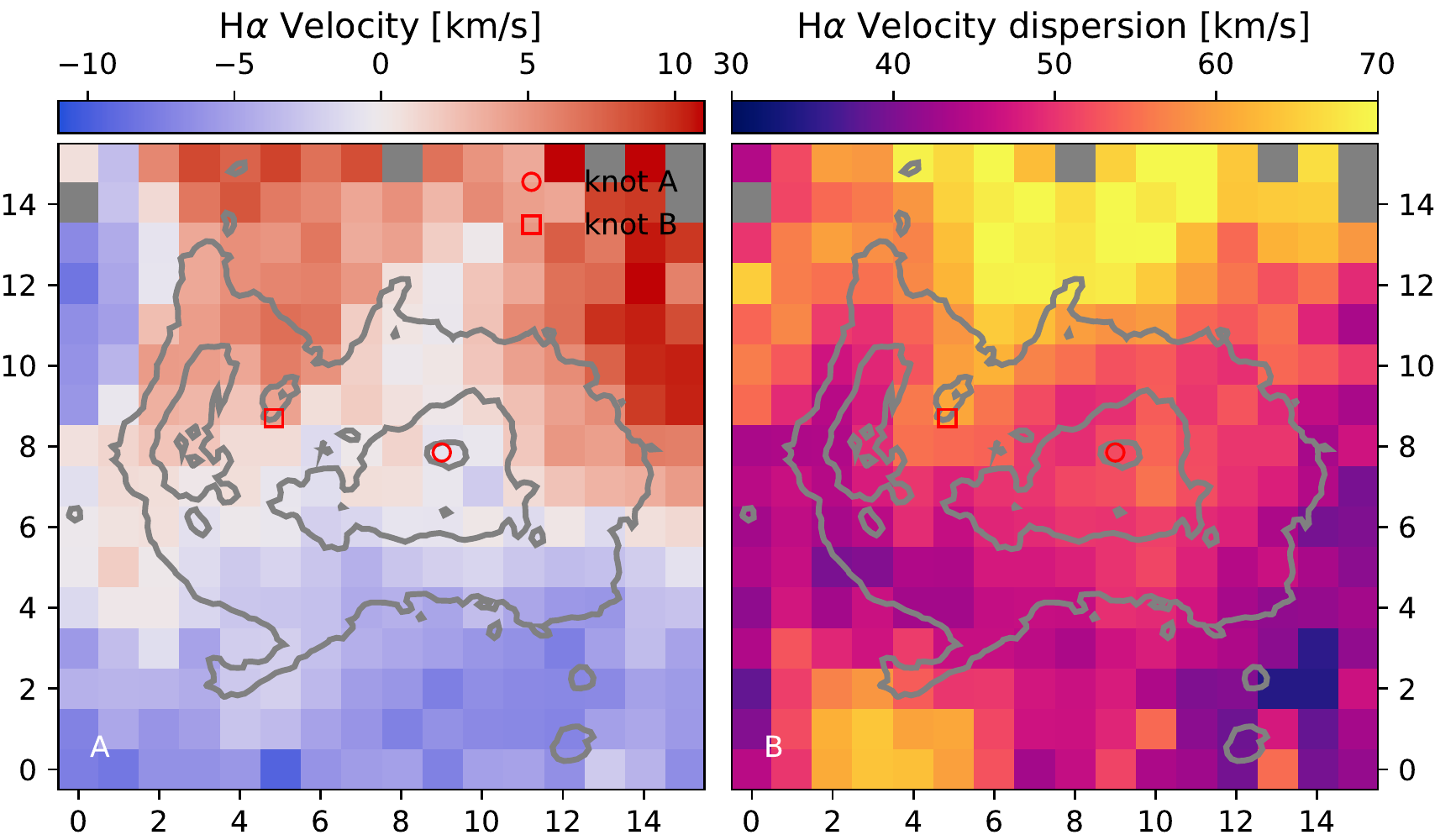}
       \caption{(A): H$\alpha$ radial velocity distribution around the  
         luminosity-weighted average velocity zero-point of $95.34$
         km/s. HST \ha\ countours are overplotted in gray solid lines, with
         the outer most (faintest) contour tracing the 
         $1.9\times10^{-7}$ erg/s/cm$^2$/\AA/arcsec$^2$ iso-intensity level.
         (B): H$\alpha$ velocity dispersion, corrected only for
         instrumental broadening. Only spaxels with
         S/N$\ge10$ are shown. The gray spaxels at positions (0,14) and
         (8,15) are dead fibers.}\protect\label{fig:veldisp} 
     \end{center}
   \end{figure*}
\section{Kinematics}\protect\label{sec:kinematics}
To investigate the kinematics, we use the H$\alpha$ line, which is the
strongest observed line in our wavelength range and has high S/N$>10$ is
all spaxels.
\subsection{Velocity field}\protect\label{sec:vel}
The radial velocity of the gas in the line of sight is obtained in the usual
way, by measuring the offsets of the line peak and comparing to its restframe
wavelength, $v=c\Delta\lambda/\lambda_\mathrm{rest}$. As a reference
zeropoint, we take the luminosity-weighted average of $\left<v\right>=95.34$
km/s. The resulting velocity map is shown in Figure \ref{fig:veldisp}a. The
location of knots A and B, as well as HST continuum-subtracted H$\alpha$
contours are overplotted for reference, with the faintest (outer most) contour
defining what we consider the ``core'' region of Mrk 71. We observe a velocity gradient, which is likely a combination of the
velocity field of the host galaxy \citep[e.g.,][]{Hunter2001} and the presence
of a biconical outflow, which we will discuss in the next section (\ref{sec:veldisp}).
A wide band of close-to-zero relative velocity, (white spaxels in Figure
\ref{fig:veldisp}a), reminiscent of a kinematical axis, runs through the
entire field of view, along an approximately east-west direction, and varies
in width between $16.5$ pc to the West, $116$ pc at the location of Knot A,
and $66$ pc to the East. Since the field of view covers only Mrk 71, which is
a part of the larger parent galaxy NGC 2366 \citet[e.g.,][]{Noeske2000}\footnote{Mrk 71 is often misidentified as NGC 2363 in the literature, when in fact NGC 2363 is an interacting dwarf galaxy to the west of Mrk 71}, the observed velocity
field is difficult to interpret in terms of lack or presence of rotation.  

To estimate the large-scale motion of the gas, we measure the shearing
velocity as $v_{\mathrm{shear}}=(v_{\rm max} - v_{\rm min})/2$. Following
\citet{Herenz2016}, we take the median of the upper and lower fifth percentile
of the velocity values to obtain $v_{\rm max}$ and $v_{\rm min}$, and
propagate the full width of each percentile to estimate conservative
uncertainties on $v_{\mathrm{shear}}$. We obtain
$v_{\mathrm{shear}}=9.6\pm3.0$ and $5.7\pm0.9$ km s$^{-1}$ for the full field
of view and for the region inside of the overplotted contours in Figure
\ref{fig:veldisp}, respectively, with measurements for additional regions
tabulated in Table \ref{tab:results0}. For such low shearing velocities to be
compatible with typical $v_{\rm max}$ of H$\alpha$ rotation curves of spiral
disks \citep[e.g.][]{Epinat2010}, the inclination would have to be close to
zero, i.e. a face-on disk. The morphology of Mrk 71 in Figure
\ref{fig:Ha_images} is, however, clearly not reminiscent of a face-on disk. As previously mentioned, Mrk 71 is part of a larger host galaxy, which may have substantially higher $v_{\mathrm{shear}}$. We deduct that Mrk 71, taken in isolation from its parent galaxy, does not
manifest a disk structure, and if such structure exists, it is not compatible
with typical local rotators. Similar inconsistencies in implied $v_{\rm max}$
were observed for the LARS sample, most of which has complex velocity fields
and too low $v_{\rm max}$ by a factor of more than two compared to typical
rotator samples \citep{Herenz2016}.  

%
%
\begin{table}
\caption{Global H$\alpha$ kinematic parameters for the full field of view
  (FoV) and for regions defined in Figure \ref{fig:spectra}.}   
\label{tab:results0}      
\centering        
\small  
\begin{tabular}{l c c c }     
\hline\hline       
Region    & $v_{\rm shear}$ & $\left<\sigma_{\rm m}\right>$ & $v_{\rm shear}/\left<\sigma_{\rm m}\right>$ \\
          & [km s$^{-1}$]  &  [km s$^{-1}$] & \\
\hline  
FoV       &  $9.6\pm3.0$ & $50.7\pm2.6$ & $0.2\pm0.06$\\
Core      &  $5.7\pm0.9$ & $50.5\pm2.9$ & $0.1\pm0.02$\\
East      &   $4.4\pm0.5$ & $48.4\pm4.5$ & $0.1\pm0.01$\\
West      &   $4.5\pm0.5$ & $51.5\pm4.2$ & $0.1\pm0.01$\\
North     &   $2.5\pm0.4$ & $51.2\pm3.6$ & $0.05\pm0.01$\\
South     &   $3.7\pm0.7$ & $46.9\pm4.5$ & $0.08\pm0.02$\\
\hline
\normalsize                  
\end{tabular}
\end{table}

\subsection{Velocity dispersion}\protect\label{sec:veldisp}
The velocity dispersion is similarly obtained from the observed broadening of the line,
$\sigma=c\mathrm{FWHM}/(2\sqrt{2\ln{2}}\lambda_\mathrm{rest})$, where FWHM is the full width at
half maximum of the line. The resulting velocity dispersion map is shown in Figure
\ref{fig:veldisp}b, corrected only for instrumental broadening. A spatially
resolved thermal broadening correction is applied in Section \ref{sec:thermal_broadening}, and does
not change the observed pattern. Two regions to the north and to the south show clearly elevated
values of $\sigma$, suggestive of the presence of an outflow. A blowout and
outflow, coincident with the location of the north region of increased
$\sigma$, has been detected via HST nebular imaging \citep[e.g.][]{James2016},
spatially resolved [OIII] kinematic data \citep[e.g.][]{Roy1991}, and H
{\small I} mapping \citep[e.g.,][]{Hunter2012}. As pointed out in
\citet{Micheva2017b}, knot B appears to be the cause of a superbubble, with 
shell morphology to the east, and the blowout region to the north. This is
consistent with the location of knot B being at the apex point of the
seemingly conical region of higher $\sigma$ to the north. Such a conical
structure is similar to that of a perturbation, propagating downstream in a
supersonic flow, as we discuss in more detail in Section
\ref{sec:mach_number}. This region also overlaps with a corridor of high [O
  {\small III}] $5007$/[O {\small II}] $3727$ ratios $\ge10$, indicating gas
with high degree of ionization, and reaching the outskirts of Mrk 71
\citep[][their figure 4]{Micheva2017b}. 

Comparing to Figure \ref{fig:veldisp}a, we note that the north and south
regions of high $\sigma$ show redshifted and blueshifted velocities,
respectively. Further, the HST continuum-subtracted H$\alpha$ contours,
overplotted on the image, show extended ``fingers'' of \ha\ emission
connecting to both high-$\sigma$ regions, and likely a product of the gas
being swept up by the outflowing gas. We cannot investigate the southern region in 
any greater detail as we are limited by the field of view. However, taken at
face value Figure \ref{fig:veldisp}b suggests that the outflow, first detected
by \citet{Roy1991}, may be biconical. The hot young stars, capable of providing
significant mechanical feedback in the form of stellar winds, as well as
expanding supernovae remnants, are likely to be located inside of Mrk 71, and
not in the outskirts, where the increased velocity dispersion $\sigma$ is
observed. Therefore the regions with higher $\sigma$ to the north and south
must be due to a sudden drop in the gas density.   

The reasoning behind this conclusion is as follows. The total velocity dispersion can be the result of both thermal and turbulent contributions. The strong anti-correlation of the temperature with gas density yields a higher thermal velocity dispersion in low-density gas. For the turbulent component we also expect smaller dispersions for higher densities. Dense regions condense out of the diffuse warm gas and consequently correspond to a contraction from large to small scales. Turbulence -- and thus the turbulent velocity dispersion -- cascades down from large to small scales with a strong inverse scaling of the dispersion as a function of density. As both thermal and turbulent velocity dispersions follow the same trend we expect an anti-correlation of $\sigma$ with the density, irrespective of the relative weighting between the two components.

\begin{figure*}
     \begin{center}
     \includegraphics[width=\hsize]{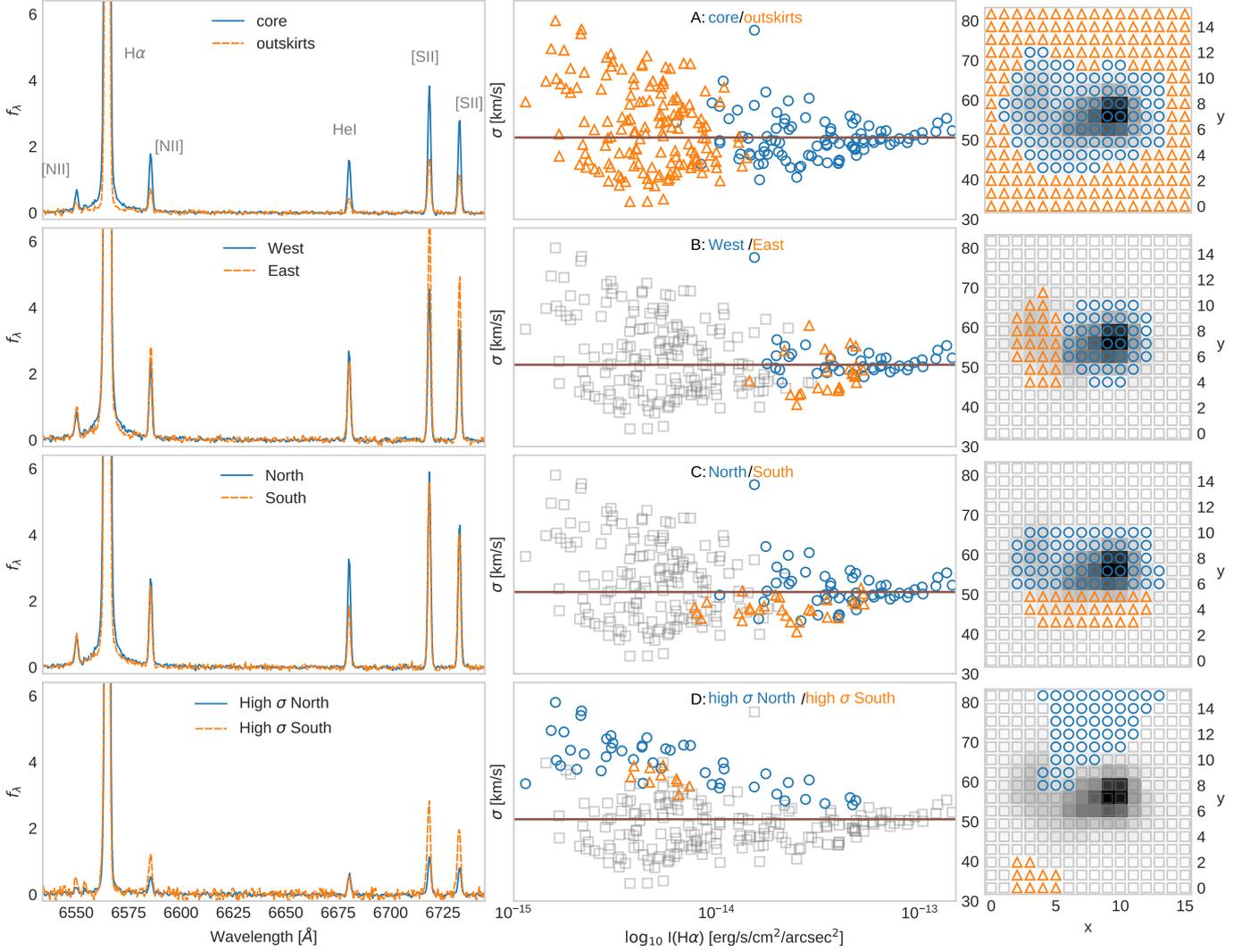}
      \caption{Left column: Averaged spectra for different spatial regions, in
        arbitrary flux density units, scaled to match in \ha\ equivalent width. The continuum has
        been subtracted from both the blue and orange spectra, enabling comparison between 
        panels in different rows. Middle column: Intensity$-$ velocity
        dispersion diagrams for \ha. The horizontal line in each panel is
        $\left<\sigma_m\right>$. Markers are color-coded by the spatial location of their corresponding spaxels, shown in the right column. Note that the y-axis values are displayed on
        the right-hand side. Right column: Spatial maps of H$\alpha$, 
        identifying the different regions. North is up, east is to the left.
      }\protect\label{fig:spectra}
      \end{center}
   \end{figure*}

In this context we can also compute the $v_{\rm shear}/\sigma_{\rm m}$ ratio,
which is an indication of whether the gas is dominated by chaotic or ordered
motion. As a representative velocity dispersion of the region we take the
intensity-weighted local mean of the resolved velocity dispersion,
$\left<\sigma_m\right>=\sum{\sigma_iI_i}/\sum{I_i}$
\citep[e.g.][]{Ostlin2001,Green2010,Glazebrook2013}. Since Mrk 71 is a giant H{\small II}
region, part of a larger parent galaxy, we expect dispersion-dominated
kinematics, and hence a ratio $v_{\rm shear}/\sigma_{\rm m}<1$. Indeed, the
tabulated values in Table \ref{tab:results0} for the full field of view, and
for regions defined in Figure \ref{fig:spectra}, are indicative of kinematics
dominated by turbulence. 

\subsection{Is there a broad component?}\protect\label{sec:broad}
A faint and very broad velocity component of FWHM$=45$\AA\ has been detected by
\citet{Roy1991}, and confirmed by \citet{GonzalezDelgado1994}, but with
FWHM$=30$ \AA. This velocity component has been reported as centered on a
cavity in the \ha\ emission, near knot A. The nature and acceleration mechanism of
this component is still not known, as there are problems with every scenario
for its origin: Thompson scattering by hot gas, superbubble blowout, stellar
winds, supernova remnants \citep[][]{Roy1992}, and turbulent mixing layers 
\citep[][]{Binette2009}. 

To facilitate our discussion, in Figure \ref{fig:spectra} we divide Mrk 71
into several regions. We aim to spatially map and identify differences in the
properties of the ionized gas, but other than knowing the location of the two
super star clusters (SSCs) knots A and B, we have no prior notion of the
extent of any regions, manifesting different properties. We therefore simply
divide Mrk 71 into the somewhat arbitrary regions of ``East'', ``West'',
``North'' and ``South''. The extent of the regions is arbitrary in the sense
that no physical criteria were applied in the selection of included spaxel,
and neighboring spaxels were excluded if, e.g., a residual cosmic ray was
present near the \ha\ line. The ``core'' region contains most spaxels inside
the outermost \ha\ contour. We also added two regions with the spaxels
covering the elevated velocity dispersion in the north and in the south. 

In Figure \ref{fig:spectra} we juxtaposition the average-stacked spectra of
these regions in the left column, plot them in an intensity $-$ velocity
dispersion diagram in the middle column, and show their spatial location in a
2D map in the right column. The color-coding is intended to facilitate quick
identification of the regions across columns. For example, in the first row
the average-stacked spectrum of the ``core'' region is shown in blue, and in
the second and third columns blue markers represent data from spaxels inside
of that region. 

Some obvious differences and features in the spectra are immediately
apparent. The shape of the \ha\ profile varies across regions, which we
examine in detail in the discussion (Section \ref{sec:discussion}). The ``West'' region has stronger
high ionization lines (He {\small I} $\lambda 6678$) and weaker [S {\small
    II}] $\lambda\lambda6716,6731$, compared to the ``East.'' This is
consistent with the higher excitation and presence of younger massive stars in
knot A compared to B \citep[e.g.][]{GonzalezDelgado1994,
  Drissen2000,James2016,Micheva2017b}. Further, high ionization gas, traced by
He {\small I} $\lambda 6678$, seems to be present also outside of the ``core''
region, as shown in row four for the high $\sigma$ north and south
regions. This is consistent with the presence of optically thin gas in the
outskirts of Mrk 71, detected via ionization parameter mapping with [O {\small
    III}] $\lambda\lambda4959,5007$ / [O {\small II}] $\lambda3727$
\citep{James2016, Micheva2017b}. 

Figure \ref{fig:spectra} illustrates that a broad component is indeed
present. The left column of the figure demonstrates that the
``core'' region has significantly broader \ha\ emission than the
``outskirts'' (row one). Dividing the core into ``East'' and ``West'' as indicated in the figure, one can further see that the \ha\ line is broader in
the west region than in the east (row two). Dividing the core into ``North'' and ``South'' regions instead decreases the difference between the \ha\ line
widths (row three), but both profiles appear broader than the ``outskirts''
region in row one. The broad component is entirely undetectable in row four 
of the figure, where the northern and southern high velocity dispersion
regions are juxtapositioned instead. Thus, the broad component is most
prominent in the core of Mrk 71, and more precisely in the western, circularly-shaped region
marked in the second row of the figure. This region encompasses knot A, but is
not centered on it. It is also apparent that the broad and narrow velocity
components are decoupled from each other, in the sense that the high velocity
dispersion northern and southern regions (row four) do not have a detectable
broad component. 

%
   \begin{figure*}
     \begin{center}
     \includegraphics[width=\hsize]{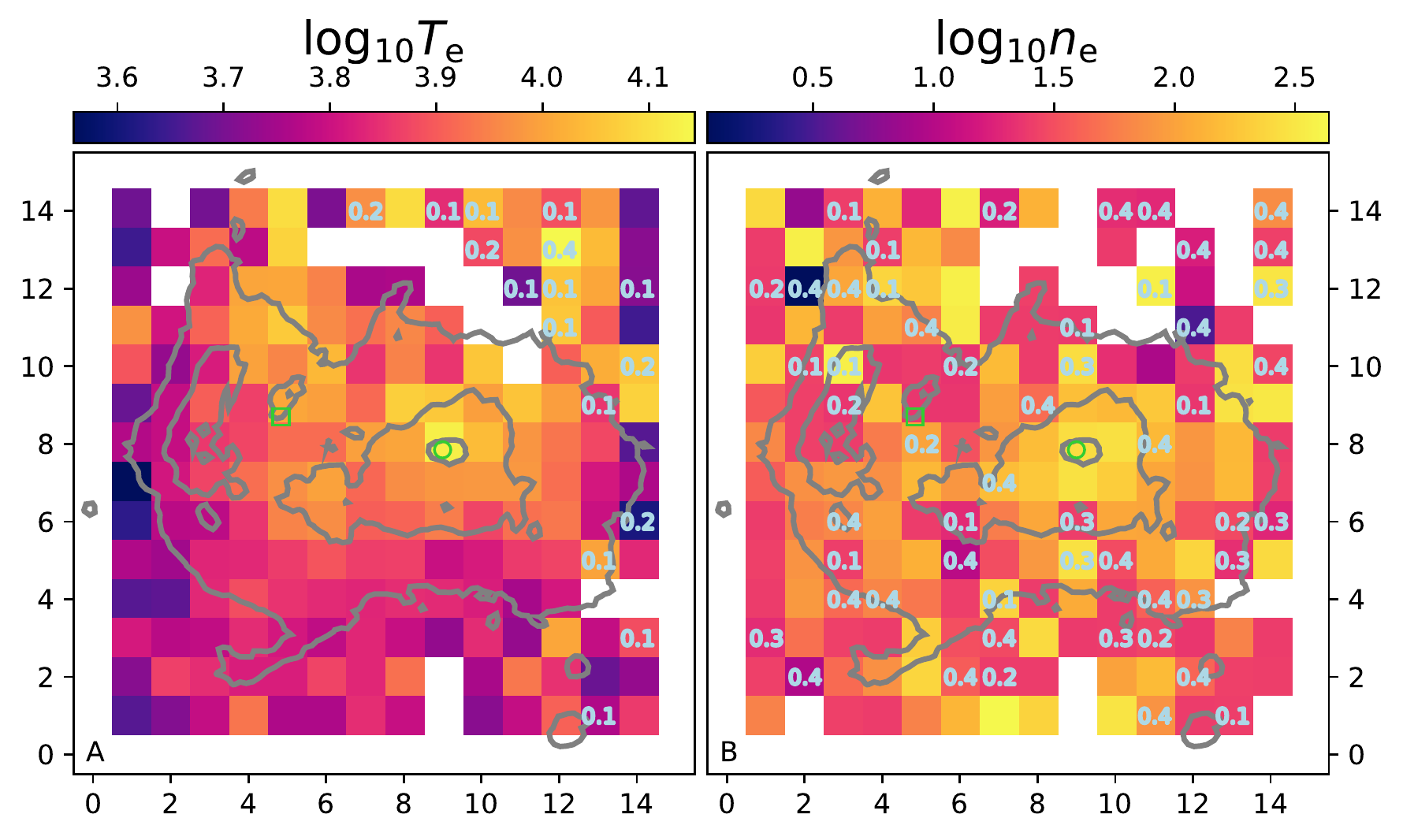}
      \caption{(A): Electron temperature $\log_{10} \rm T_{e} [K]$. Knots A
        (green open circle) and B (green open square), as well as
        \ha\ contours are overplotted for orientation. Numbers indicate
        relative error, with '0.1' corresponding to
        $0.1\leq\delta\mathrm{T_e/T_e}<0.2$, etc. Spaxels with no numbers have 
        $\delta\mathrm{T_e/T_e}<0.1$. (B): Electron density $\log_{10} \rm n_{e} 
        [cm^{-3}]$; same legend as in (A). Both $\rm T_{e}$ and $\rm n_{e}$
        are obtained with \texttt{pyneb}.
        }\protect\label{fig:temden}
      \end{center}
   \end{figure*}
\subsection{The I$\mbox{-}\sigma$ diagnostic}\protect\label{sec:Isigma}
\citet{MunozTunon1996} suggest that an intensity versus velocity dispersion
diagram can help identify the main line broadening mechanism, caused by, e.g.,
expanding shells and loops generated by the winds of massive stars. The I$\mbox{-}\sigma$ diagnostic uses the fact that when fitting line profiles with a single Gaussian regardless of the shape of the line, the fit will conserve the flux. As a consequence, strongly-split and very asymmetric line profiles will result in a fit of lower peak intensity but broader shape. This can be used to identify line fits with low intensity and broad shapes. If spatially clustered together, such points suggest a region of very asymmetric and/or split emission lines, which are usually the signatures of the expansion of the gas in shells, rings, bubbles. In the I$\mbox{-}\sigma$ parameter space such regions will appear as ``inclined bands'', because the broader the line appears, the lower its peak flux. We construct such a diagram in the middle column of Figure \ref{fig:spectra}, but
we use the modified presentation of \citet{Moiseev2012} with $\log{I}$ being
the integrated flux of the line, and not the peak intensity. This has the
advantage of being independent of the spectral resolution. Here we use the
velocity dispersion $\sigma$, shown in Figure \ref{fig:veldisp}b, i.e.,
corrected only for instrumental but not thermal broadening. The horizontal
line in these plots is the intensity-weighted local mean velocity dispersion,
$\left<\sigma_m\right>$, defined in Section \ref{sec:veldisp}.

In this diagram, we see the typical horizontal lane of
$\sigma\approx\left<\sigma_m\right>$, made up of the majority of spaxels in
the ``core'' region. In our case, this $\left<\sigma_m\right>=50.7$ km/s
corresponds, within errors, to the minimum line width, observed in the
``core'' region (Figure \ref{fig:spectra}A). Such behavior is consistent with
the \citet{MunozTunon1996} interpretation of this diagnostic. This value
should be close to the stellar velocity dispersion, $\sigma_\star$. In
addition to the horizontal lane, we observe lower-intensity points with both
higher and lower $\sigma$. According to the model schematics of \citet[][their
  figure 6]{MunozTunon1996}, this implies that the expanding shells and loops
around massive stars have breached the ``kinematic'' core of Mrk 71, and are
expanding beyond its borders. The spatial map of the velocity dispersion, with
the overplotted \ha\ contours in Figure \ref{fig:veldisp}b, is clearly
consistent with such a scenario. While it is difficult to identify new,
previously undetected, expanding shells or loops with this diagnostic at our
course spatial resolution, we see that the high $\sigma$ north and south
regions (row four), indeed appear as inclined bands in this diagram. At least
the northern region has been previously identified as a blowout region by,
e.g., \citet[][]{Roy1991}. 

Finally, in the model of \citet{MunozTunon1996}, the smaller scatter of the west region
points in Figure \ref{fig:spectra}B suggests that the ``West'' region, hosting
knot A, is dynamically younger than the ``East'' region, hosting knot B. This
is consistent with the younger age of knot A compared to knot B.

%
%
\begin{table*}
\caption{Statistics in T$_e$, n$_e$,
  $\sigma_{\mathrm {thermal}}$, $c_S$, and $\mathcal{M}$ for the regions
  defined in Figure \ref{fig:spectra}. For each quantity, columns (1), (2), and (3), are the average (rms), median, and the brightest spaxel values, respectively.}   
\label{tab:results1}      
\centering        
\small  
\begin{tabular}{l c c c c c c c c c c c c c c c c }     
\hline\hline       
          &\multicolumn{3}{c}{$\rm T_{e}$ [K]} & \multicolumn{3}{c}{$\rm n_{e}$ [cm$^{-3}$]} & \multicolumn{3}{c}{$\rm \sigma_{thermal}$ [km s$^{-1}$]} & \multicolumn{3}{c}{$\rm c_{S}$ [km s$^{-1}$]} &   \multicolumn{3}{c}{$\mathcal{M}$} \\ 
  Region  & (1) & (2)  & (3)    & (1) & (2)  & (3)    & (1) & (2)  & (3)    & (1) & (2)  & (3)    & (1) & (2)  & (3)    \\
\hline  
Core      &  8140(90) & 7930 &  13384 & 101(10) &  77 & 371 & 8.1(2.9) & 8.1 & 10.5 &  11.5(3.4) & 11.4 &  14.9 &  4.3(2.1) & 4.3 &  6.0  \\ 
Outskirts &  7239(85) & 6639 &  13916 & 89(9)   &  27 & 437 & 7.6(2.8) & 7.4 & 10.7 &  10.8(3.3) & 10.5 &  15.2 &  5.0(2.2) & 4.9 &  7.9  \\ 
East      &  7620(87) & 7310 &  10302 & 79(9)   &  64 & 371 & 7.9(2.8) & 7.8 & 9.2  &  11.2(3.3) & 11.0 &  13.0 &  4.3(2.1) & 4.3 &  5.0  \\ 
West      &  8925(94) & 8707 &  13384 & 110(11) &  93 & 290 & 8.5(2.9) & 8.5 & 10.5 &  12.1(3.5) & 12.0 &  14.9 &  4.2(2.0) & 4.2 &  5.2  \\ 
North     &  8723(93) & 8608 &  13384 & 99(10)  &  81 & 371 & 8.4(2.9) & 8.4 & 10.5 &  11.9(3.5) & 11.9 &  14.9 &  4.2(2.1) & 4.2 &  5.2  \\ 
South     &  6780(82) & 6889 &  7839  & 85(9)   &  42 & 290 & 7.5(2.7) & 7.5 & 8.0  &  10.6(3.3) & 10.7 &  11.4 &  4.3(2.1) & 4.3 &  4.9  \\ 
\hline
\normalsize                  
\end{tabular}
\end{table*}

\section{Physical conditions}\protect\label{sec:physcond}
\subsection{Electron temperature and density}\protect\label{sec:temden}
The density-sensitive [S {\small II}] ratio can be used to map the electron
density at each spaxel. The fortuitous presence of a second order spectrum of
the UV [O {\small II}] doublet enables us to also calculate an electron
temperature map using the temperature-sensitive [O {\small} II]
$\lambda3726+3729/$[O {\small II}] $\lambda 7320+7330$ ratio. Using the python
software \texttt{pyneb} \citep{Luridiana2015}, we 
cross-converge the temperature (T$_e$) and density (n$_e$) from both
diagnostics with \texttt{getCrossTemDen}. To obtain robust estimators for
these quantities and their uncertainties, we perform $1000$ Monte Carlo runs,
at each realization perturbing the spectra in the manner described in Section
\ref{sec:linemeasures}, and obtaining a new T$_e$ and n$_e$ after re-measuring
the line ratios. The average T$_e$ and n$_e$ of these realizations are
presented in Figure \ref{fig:temden}, and listed in tables \ref{tab:results1}
and \ref{tab:results2} for different regions. Due to the non-parametric nature of the
resulting distributions, the associated uncertainties in each spaxel are
estimated as the root mean square (rms) of the $1000$ T$_e$ and n$_e$
values. For the case of the electron density, spaxels in the low-density limit
but inside of the high signal-to-noise (S/N) region, indicated by the
overplotted contours in the figure, are explicitly set to the lowest physical
value of [S {\small II}]$\lambda6731/6716\sim0.69$.

%
   \begin{figure*}
     \begin{center}
       \includegraphics[width=\hsize]{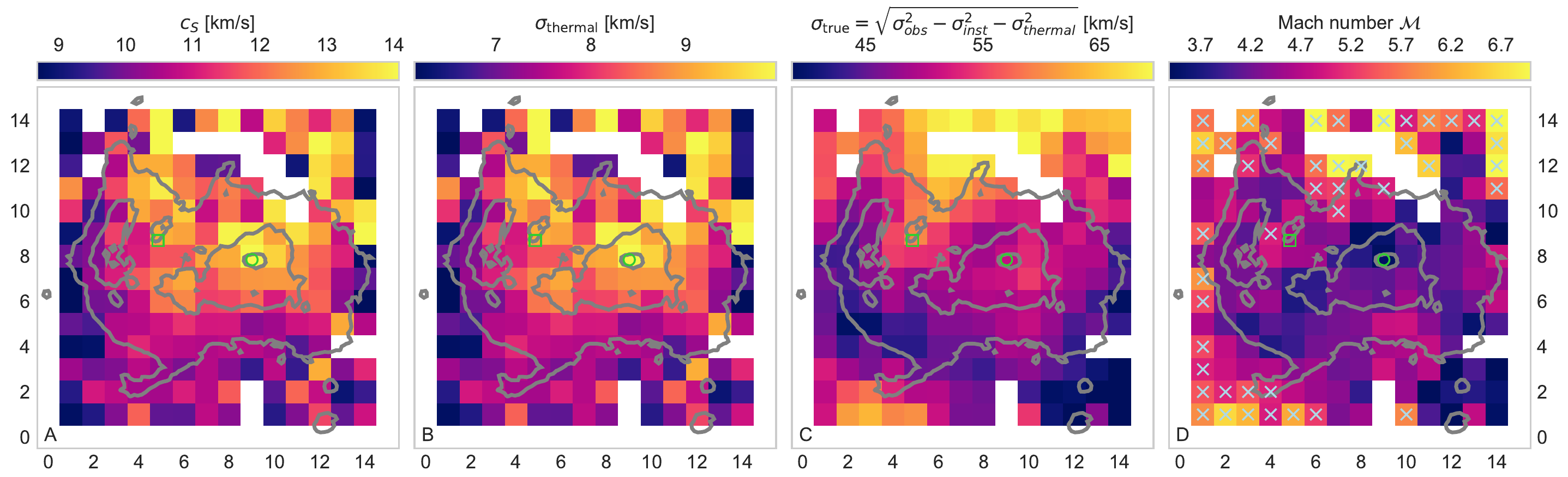}
       \caption{(A): Speed of sound $c_S$. Knots A (open circle) and B (open
         square) are given for orientation. (B): Thermal broadening. (C)
         ``True'' velocity dispersion, corrected for both instrumental and
         thermal broadening. (D) Mach number $\mathcal{M}$ map, with strictly
         hypersonic regions ($\mathcal{M}\ge5.0$) marked with 'x'. At knot B,
         $\mathcal{M}=4.63$.}\protect\label{fig:sound}   
     \end{center}
   \end{figure*}

The temperature-sensitive [O {\small} II] $\lambda3726+3729/$[O {\small II}]
$\lambda 7320+7330$ ratio spans a very wide wavelength range, and so is
dependent on how accurate the internal extinction can be estimated. In Section
\ref{sec:redcor} we obtained \ha/\hb\ from HST imaging, using a constant scaling
factor to subtract the continuum in each filter. While our reddening map is
consistent with the literature, it is worth considering the effects on the
electron temperature that under- or overestimating the extinction correction c(\hb)
would have. If one underestimates c(\hb), the UV [O {\small II}] flux will be more strongly
affected than the [O {\small II}] doublet at $7325+$\AA, the
temperature-sensitive [O {\small II}] ratio will be underestimated, and hence
the electron temperature will be overestimated.

%
%
\begin{table}
\caption{Average, median and brightest spaxel differences in T$_e$, n$_e$,
  $\sigma_\mathrm{thermal}$, $c_S$, and $\mathcal{M}$ between the regions
  defined in Figure \ref{fig:spectra}.}             
\label{tab:results2}      
\centering          
\begin{tabular}{c c c c}     
\hline\hline       
        & Core$-$Outskirts & East$-$West & North$-$South \\\hline
        & \multicolumn{3}{c}{$\Delta$ T$_e$ [K]}\\
Average & 1114           & -1343     & 1751        \\ 
Median  & 1202           & -1457     & 1717        \\ 
Max     & 292            & -3077     & 5524        \\ \hline
        & \multicolumn{3}{c}{$\Delta$ n$_e$ [cm$^{-3}$]}\\
Average & 9              & -29       & 10         \\ 
Median  & 48             & -42       & 46         \\ 
Max     & 45             & 160       & 257        \\ \hline
        & \multicolumn{3}{c}{$\Delta \sigma_\mathrm{thermal}$ [km/s]}\\
Average & 0.6            & -0.7       & 0.9         \\ 
Median  & 0.7            & -0.7       & 0.9         \\ 
Max     & 0.1            & -2.5       & 1.3         \\ \hline
        & \multicolumn{3}{c}{$\Delta$ c$_s$ [km/s]}\\
Average & 0.8            & -1.0       & 1.3         \\ 
Median  & 0.9            & -1.3       & 1.0         \\ 
Max     & 0.2            & -1.8       & 3.5         \\ \hline
        & \multicolumn{3}{c}{$\Delta \mathcal{M}$}\\
Average & -0.6            & 0.1       & -0.1         \\ 
Median  & -0.6            & 0.1       & -0.05         \\ 
Max     & -1.4            & 0.2      & 0.4         \\ 
\hline                  
\end{tabular}
\end{table}

We therefore compare our T$_e$ and n$_e$ maps with results from long-slit spectra
in literature. For the spaxel containing knot A, T$_e=13 374\pm116$ K, and
n$_e=273\pm17$ cm$^{-3}$. Similar values obtained from the same line ratios
are found in the literature, suggesting that our unusual procedure for
obtaining the T$_e$ map has nevertheless been successful. For knot A,
\citet{Sokal2016} measure T$_e$([O {\small II}])$=1.3\pm0.3\times10^4$ K, and
\citet{GonzalezDelgado1994} obtain T$_e$([O {\small
    II}])$=1.45\pm0.13\times10^4$ K, both consistent with our measurement
within the uncertainties. Similarly, n$_e=512\pm264$ cm$^{-3}$ in
\citet{Sokal2016}, and n$_e=235\pm41$ cm$^{-3}$ in
\citet{GonzalezDelgado1994}, also both consistent with our measurement. Note
that we also obtained an electron density map from the UV [O {\small
    II}]$3726/3729$ ratio, which gave n$_e=287$ cm$^{-3}$ for knot A -- a
result consistent with that obtained from the [S {\small II}] ratio.

The electron density map is quite noisy. A noticable increase in n$_e$ is
observed around knot A, which is likely real, given its location (knot A is an
embedded super star cluster). The spaxels with similarly high n$_e$ found in
the outskirts of the image, however, are more reminicent of salt-and-pepper
noise, than real features. Taken at face value, the electron density map shows
densities typical for H {\small II} regions over the entire field of view.

As can be seen in Figure \ref{fig:temden}, the dominant trend in T$_e$
presents a vertical pattern, with T$_e$ being significantly higher to the
North around knot A than in the South. The
observed change in temperature across Mrk 71 in the figure may be due to
metallicity variations on the same spatial scales -- a scenario investigated by \citet{James2016}. We are unable to verify
this, as none of the [O {\small III}] lines are in our spectral coverage, and
therefore we cannot use the so-called ``direct'' T$_e$ method to determine the
oxygen abundance. Additionally, \citet{Kewley2002} demonstrate that
metallicity-sensitive strong-line method ratios like [N {\small II}]/[O
  {\small II}], [N \small{II}]/[S {\small II}], or [N {\small II}]/H$\alpha$,
\normalsize all within the spectral range of our data, become insensitive to
the metallicity at $Z\lesssim0.5$Z$_\odot$. Since Mrk 71 is metal-poor
\citep[$12+\log{\mathrm{O/H}}=7.89\pm0.01$;][]{Izotov1997,Hunter1999}, the
``strong-line'' method involving these lines also cannot be used to obtain spatially resolved
metallicity.

However, at a projected separation of $5$ arcsec, the two knots A and B have very
similar oxygen abundances of $12+\log{\mathrm{O/H}}=7.89\pm0.01$ and
$7.85\pm0.02$, respectively, measured using the ``direct'' method on
data from long slit spectra \citep{Izotov1997}. \citet{GonzalezDelgado1994}
also find no significant variation in the oxygen abundances from spectra
extracted from $15$ positions, including knots A and B, from long slit
observations. This is expected, since they are also unable to find significant T$_e$
gradients for these positions, because their temperature uncertainties are
quite large and their slit positions generally avoid the low-temperature
regions. Further, \citet{James2016} use HST narrowband imaging and the 
$\mathrm{R_{23}}$ index (``strong-line'' method) to obtain a map of
$12+\log{\mathrm{O/H}}$ in a $5\arcsec\times5\arcsec$ region, covering knot
A. Within the uncertainties of their measurement there is no convincing trend
in the spatial metallicity distribution.  Although discrepancies in
temperature between the ``strong-line'' and ``direct'' T$_e$ methods are
expected, relative variations would be traced by both. Therefore, the observed
seeming variation in the electron temperature may reflect true differences in
the physical conditions of the gas. The highest observed T$_e$ coincides with
the location of knot A, where numerous evidence points to the possible
existence of very massive stars \citep[e.g.,][]{James2016,Micheva2017b}. The high T$_e$ is likely the result of the very high ionization parameter of log U$=-1.89\pm0.3$ and the associated extremely high ratio of [O {\small III}]$\lambda5007$/[O {\small II}]$\lambda3727=23.0$ at knot A \citep{Micheva2017b}. This is consistent with the observed elevated T$_e$ in the north (Figure \ref{fig:temden}a), in a region which also shows high [O {\small III}]$\lambda5007$/[O {\small II}]$\lambda3727$ ratios \citep{James2016}.

Nevertheless, none of these arguments convincingly demonstrate that there is no
metallicity variation across Mrk 71, giving rise to the observed seeming variation in electron temperature. In particular, the ``South'' region,
defined in Figure \ref{fig:spectra}, is not covered by either the Izotov et
al. or the James et al. metallicity measurements, and only partially covered
by the Gonzalez-Delgado et al. data.



\subsection{Speed of sound}\protect\label{sec:speedofsound}
The electron temperature map enables us to spatially map the speed of sound
across the field of view, given by $c_S = \sqrt{\mathrm{\gamma k_BT_e/\mu m_H}}$ \citep{Osterbrock1989}. This
is the speed at which pressure waves travel in a hydrodynamical expansion of
the region. We assume a completely ionized, isothermal, pure hydrogen
\hii\ region with adiabatic index $\gamma=1$ and mean molecular weight
$\mu\approx0.5$. The resulting sound speed is largest at knot A ($c_S=14.9$ 
km/s), with average, median, and brightest spaxel differences presented in Table
\ref{tab:results2} for the regions defined in Figure \ref{fig:spectra}. The
spatial variation of $c_S$ is presented in Figure \ref{fig:sound}a, and is in the range
$7.6$ to $14.9$ km/s, with an average of $\left<c_S\right>=10.5\pm1.4$ km/s from all
spaxels in the field of view. This is a typical value for \hii\ regions,
at temperatures of $T_e\sim10^4$ K. The statistics of $c_S$ for different
regions are provided in tables \ref{tab:results1} and \ref{tab:results2}.

\subsection{Thermal broadening}\protect\label{sec:thermal_broadening}
Similarly to the speed of sound, one can obtain the thermal broadening of the
emission lines, caused by the thermal motion of the gas,
$\mathrm{\sigma_{thermal}=\sqrt{\mathrm{k_BT_e/m_H}}}$. The thermal broadening
ranges from $10.5$ km/s at knot A, to $5.4$ km/s, with an average across the
entire field of view of $\mathrm {\left<\sigma_{thermal}\right>=7.9\pm1.0}$
km/s. A map of the spatial distribution of $\mathrm {\sigma_{thermal}}$ is
shown in Figure \ref{fig:sound}b, with differences between regions tabulated
in Table \ref{tab:results2}. Compared to Figure \ref{fig:veldisp}b, some
spaxels are missing data, but the overall pattern of increased velocity
dispersion to the north and south is preserved. The missing data are due to the
propagated missing data in the thermal broadening map, which in turn is due to
the electron temperature map. The statistics of $\mathrm {\sigma_{thermal}}$ for different
regions are provided in tables \ref{tab:results1} and \ref{tab:results2}.  

\subsection{Mach number}\protect\label{sec:mach_number}
Explicitly obtaining the speed of sound and the thermal broadening allows one
to map the Mach number $\mathrm {\mathcal{M}=\sigma_{true}/c_s}$, which is simply a
measure of how fast the ionized gas is moving relative to the speed of
sound. The ``true'' velocity dispersion, shown in Figure \ref{fig:sound}c, is obtained as 
$\mathrm{\sigma_{true}^2=\sigma_{obs}^2-\sigma_{inst}^2-\sigma_{thermal}^2}$,
where $\mathrm{\sigma_{obs}}$, $\mathrm{\sigma_{inst}}$, and
$\mathrm{\sigma_{thermal}}$ are the observed, instrumental, and thermal line
broadening, respectively. The spatial distribution of $\mathcal{M}$, shown in
Figure \ref{fig:sound}d, has a range of $3.2<\mathcal{M}<7.5$, with average
$\left<\mathcal{M}\right>=4.6\pm0.7$, and hence the gas we trace is supersonic
($\mathcal{M}>1$) over the entire field of view of $16\times16$ arcsec
$=112\times112$ parsec. The statistics of the Mach number for different
regions are provided in tables \ref{tab:results1} and \ref{tab:results2}.  

\section{Discussion}\protect\label{sec:discussion}
In this section we use the nomenclature for different spatial regions, defined in Figure
\ref{fig:spectra}, namely ``East'', ``West'', ``core'', and we
refer to the ``High $\sigma$ North'' region simply as a region of increased
velocity dispersion. 

NGC 2366, dominated by Mrk 71 in terms of star formation rate density and
surface density of ionising photons, is the nearest GP analog and a candidate for
LyC escape. At its distance of $3.4$ Mpc, this galaxy offers the opportunity
to analyse possible mechanisms and channels of LyC escape. 

The proposed mechanism of LyC escape in NCG 2366/Mrk 71 involves the presence of optically
thin corridors of lower gas density, through which LyC photons can reach the
intergalactic medium \citep[][]{Micheva2017b}. In this scenario, a SSC from an initial star-forming  
event (knot B in the case of Mrk 71), powers large-scale outflows via
mechanical feedback. A consecutive star-forming event produces additional SSCs
(knot A in Mrk 71), which ionize the remaining gas in the pre-cleared region,
in the process possibly carving optically thin corridors reaching beyond the
outskirts of the galaxy. Knot B (age $\gtrsim5$ Myr), has a population of
Wolf-Rayet stars \citep{Drissen2000,Sokal2016}, whose strong stellar winds are
capable of providing substantial mechanical feedback to pre-clear the
region. Knot A is an extremely young ($\lesssim1$ Myr) SSCs, still embedded in its
natal cloud, and it dominates the ionization budget of Mrk 71
\citep{Drissen2000}. Very massive stars, younger than $\lesssim1$ Myr and with
initial masses M$_{\star}\ge100$ M$_{\odot}$ \citep[VMS;
  e.g.,][]{Crowther2010,Grafener2015}, may be present in Knot A
\citep[][]{James2016,Micheva2017b}.

%
   \begin{figure}
     \begin{center}
       \includegraphics[width=\hsize]{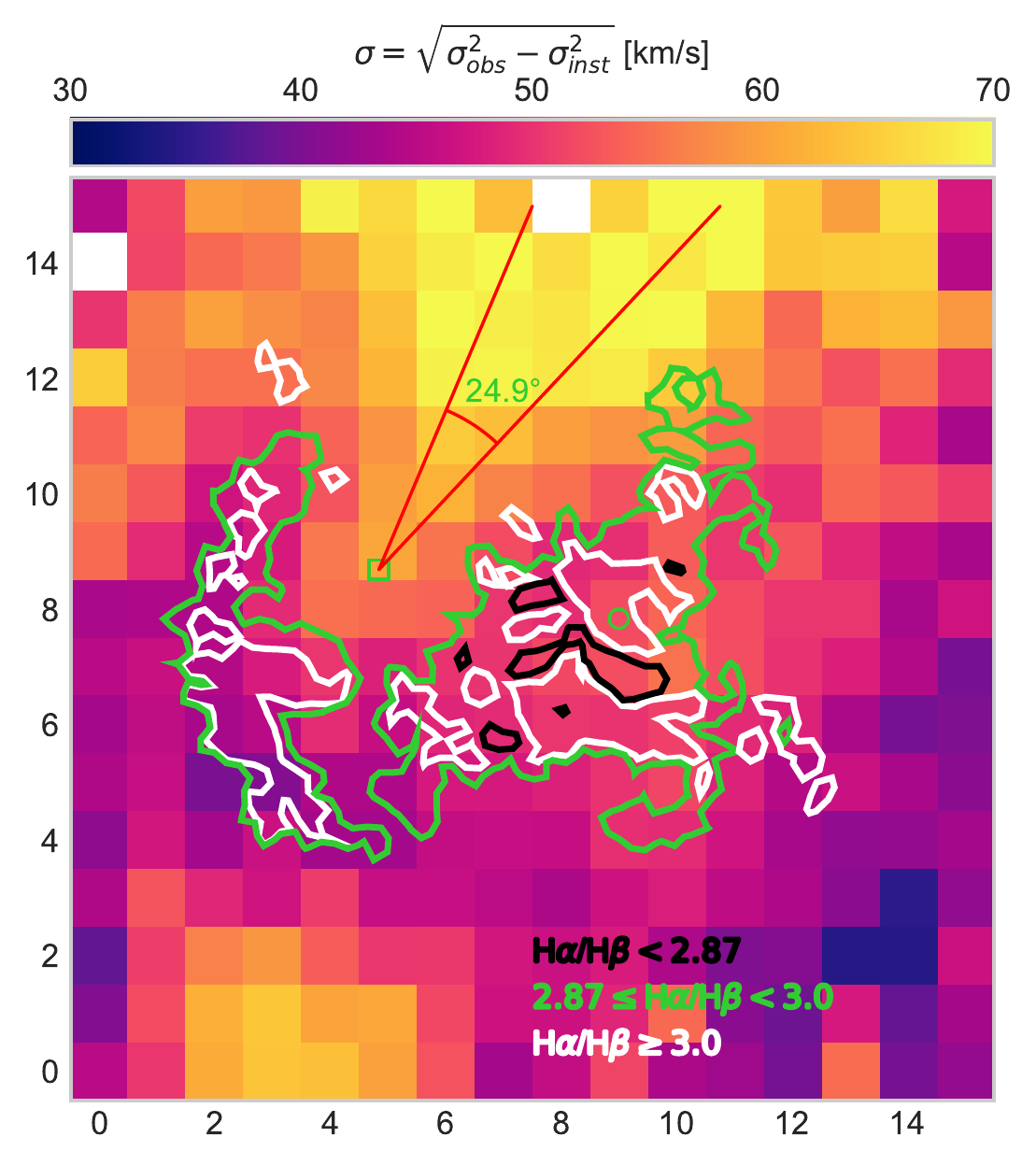}
       \caption{Velocity dispersion, corrected only for instrumental
         broadening, with overplotted dust contours. The cone of $2\times$ the
         Mach angle of downstream propagation of perturbations is also plotted
       for the Mach number at knot B (open square), $\mathcal{M}=4.63$. The
       extinction outside of the plotted contours is negligible, see the
       reddening map in Figure \ref{fig:dustmap}.}\protect\label{fig:machangle} 
     \end{center}
   \end{figure}

Neutral gas, which readily absorbs LyC photons, has been observed in NGC 2366,
with line of sight N(H {\small I}) column densities of a few $\times10^{21}$
cm$^{-2}$ at the projected position of Mrk 71 \citep{Hunter2001}, which is
prohibitively optically thick for the escape of LyC radiation. It is, however,
unlikely that all of the observed H {\small I} gas lies in front of Mrk
71. Optically thin Si {\small II} $\lambda1260/\lambda1526$ ratios, weak
detections or non-detections of other low-ionization and neutral species, and
low reddening in Mrk 71 \citep{Micheva2017b}, all suggest much lower absorption
columns, and hence that most of the H {\small I} gas is behind Mrk
71. Additionally, ionization parameter mapping of Mrk 71 \citep[][]{James2016}
shows high [O {\small III}]/[O {\small II}] ratios over most of the region,
and stretching from the core to the north beyond the outskirts of Mrk 71
\citep{Micheva2017b}. Such ratios may also indicate optically thin gas
\citep{Jaskot2013}. Supporting this scenario is the spatially coincident
outflow of ionized gas, detected with Fabry-Perot data \citep{Roy1991}, and
also visible through an increase in velocity dispersion in our Figure
\ref{fig:veldisp}b.  

Further evidence can be compiled from our own data. The region of observed
increased velocity dispersion in Figure \ref{fig:veldisp}b implies lower gas
density. Additionally, the spatially resolved Mach number map in Figure
\ref{fig:sound}d shows that while all spaxels indicate supersonic gas 
($\mathcal{M}>1$), some spaxels even show hypersonic gas
($\mathcal{M}\ge5$). These are almost entirely found in the outskirts of Mrk
71, outside of the overplotted \ha\ contours of the ``core'' region. This
boost in the relative gas velocity is a further indication of a drop in the
gas density outside of the ``core.''

Inside of the denser ``core'' region, the observed spatial distribution of the
velocity dispersion allows one to identify knot B as the likely source of mechanical
feedback which has cleared a channel of lower gas density reaching the
outskirts. In a supersonic or hypersonic flow, a perturbation will generate sound waves,
which propagate downstream and accumulate on a cone \citep{Landau1987}, whose
surface is refered to as the shock front. A cone structure is 
visible in the velocity dispersion map in Figure \ref{fig:veldisp}b, with knot
B seemingly located at the apex of the cone. The
aperture angle of the cone is $2\alpha$, where $\alpha$ is the Mach angle, 
$\sin{\alpha}=c_S/\sigma=1/\mathcal{M}$. At knot B $\mathcal{M}=4.63$, and
hence $\alpha\approx12.5\deg$. Our spatial resolution prevents us from
accurately tracing such a narrow cone. In Figure \ref{fig:machangle} we have
plotted the resulting cone aperture of $2\alpha$, approximately in the
direction of the increased velocity dispersion. The velocity dispersion in
this figure has been corrected only for instrumental broadening, in order to
display as many spaxels with data as possible. We note, that the cone falls in
the region of very low dust attenuation, which is shown by the overplotted
\ha/\hb\ contours. The seemingly biconical nature of the outflow, discussed in Section
\ref{sec:veldisp}, supports the scenario of mechanical feedback from knot B
being the cause of the low density channel, instead of the turbulent gas
simply taking a pre-existing path of least resistance. In the latter case a
biconical structure, whose axis crosses the location of knot B, is less likely,
although not impossible considering projection effects. 

   \begin{figure*}
     \begin{center}
     \includegraphics[width=\hsize]{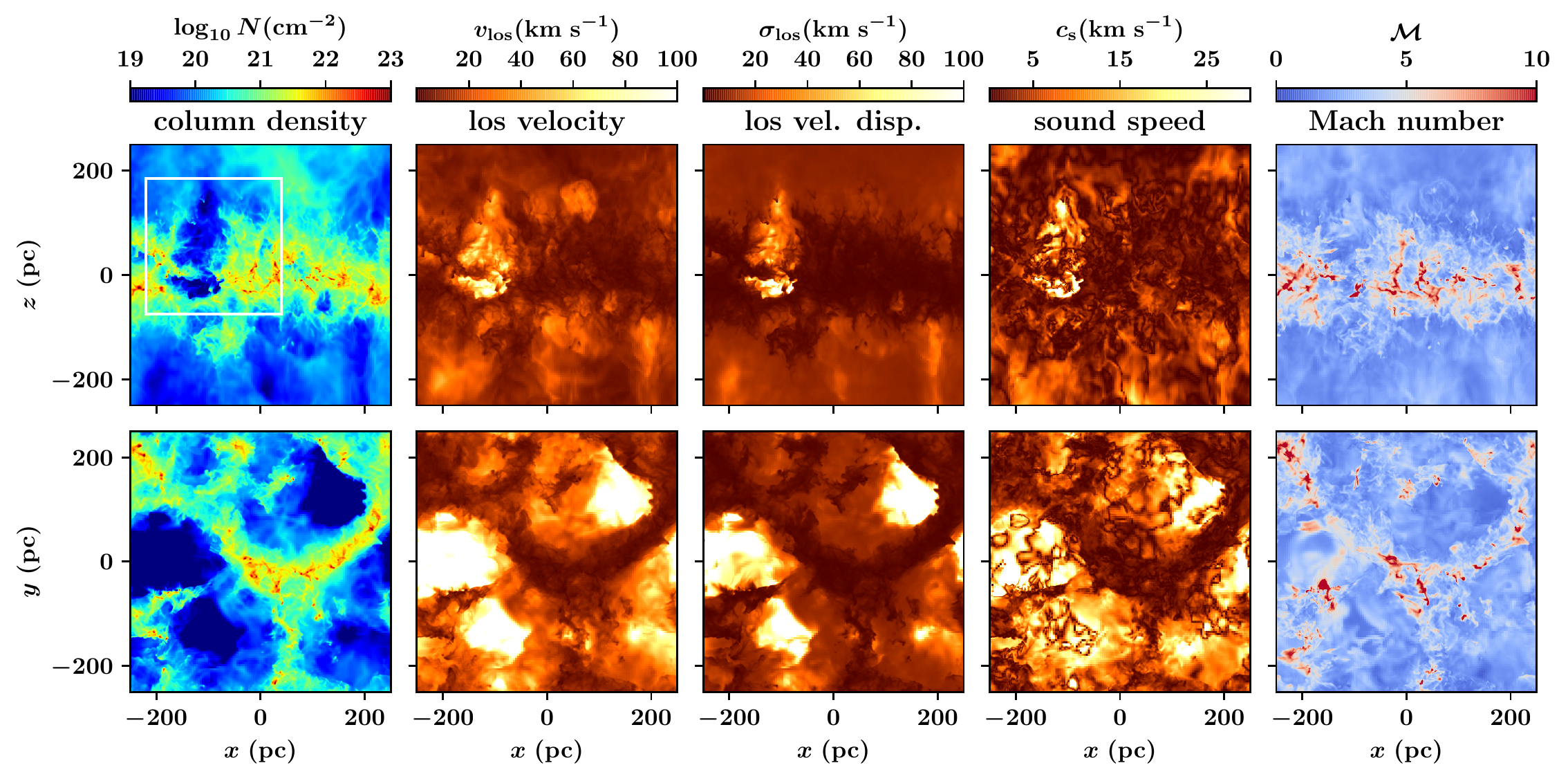}
      \caption{High-resolution runs of the SILCC simulations
        \citep{Girichidis2018}. The white rectangle indicates the area
        comparable to our observations. The top row shows edge-on projection
        of the galactic disk, the bottom one the face-on view. See Section
        \ref{sec:discussion} for details. 
        }\protect\label{fig:models}
      \end{center}
   \end{figure*}

Since the Mach number is the ratio between ``true'' velocity dispersion and
the speed of sound, both of which increase in turbulent, ionized regions, a
wide range of Mach numbers can be obtained at a given gas column density
\citep[e.g.,][their figure 13]{Girichidis2018}. To estimate if the drop in
gas density could be significant, we therefore compare our observations to
available hydrodynamical simulations of the interstellar medium. 

We use the recent high-resolution runs \citep{Girichidis2018} of the
SILCC\footnote{SImulating the LifeCycle of molecular Clouds} 
simulations \citep[][]{Walch2015,Girichidis2016} for the comparison. The setup
covers a fraction of a galactic disc in a volume of
$0.5\times0.5\times0.5\,\mathrm{kpc}^3$ at a resolution of
$1\,\mathrm{pc}$. The magneto-hydrodynamic simulations follow the thermal
state of the interstellar medium using a chemical network including ionised,
atomic and molecular hydrogen as well as C$^+$ and CO with solar values for
the metallicity. Radiation from stars is included as a backgroud radiation
field, which is locally attenuated in regions of efficient shielding. Further
stellar feedback is modelled via supernovae, where we distinguish between
individual ($1/3$) and clustered ($2/3$) explosions. The stellar component is
not modelled as individual stars. Instead we assume a stellar initial mass
function \citep{Chabrier2003} and compute an effective star formation and
resulting supernova rate based on the Kennicutt-Schmidt relation
\citep{Schmidt1959,Kennicutt1998}. The dynamical features in the simulations
are thus produced by the thermal energy injection corresponding to the
supernovae. This is surely a simplification of the stellar feedback
description. \citet{Gatto2017} and \citet{Peters2017} follow the dynamical
formation of star clusters in dense and collapsing regions and include stellar
winds and ionizing radiation from the star clusters with a maximum resolution
of 4 pc. However, focusing on the global dynamics, the general conclusions of the comparison between
simulations and observations is not altered by the additional physics in the
latter two studies and we chose the simulations by \citet{Girichidis2018}
because of the four times higher resolution.  

We chose these simulations because of the similar column density and global turbulent dynamics, i.e. similar range of measured velocity dispersion. Global galactic properties like the halo mass and the resulting dark matter distribution are not relevant for the ISM dynamics on scales of $\sim100\,\mathrm{pc}$, which we are here investigating and comparing. Gravitational effects due to stellar masses might be comparable to the effects of self-gravity of the gas. The total mass of $\gtrsim1.6\times10^5\,\mathrm{M}_\odot$ in Mrk 71 \citep{Micheva2017b} is comparable to the mass of a few $10^5\,\mathrm{M}_\odot$ in the selected region of the simulation.

The two main hypotheses in the interpretation of the observations we want to
test, are (1)  whether the physical properties of the conical outflow region are
consistent with a significant enough decrease in density to enable LyC escape,
and (2) whether the orientation of the conical outflow is along the plane of the disk,
which would likely prohibit LyC escape into the IGM, regardless of density
drop along the cone. 

   \begin{figure*}
     \begin{center}
     \includegraphics[width=\hsize]{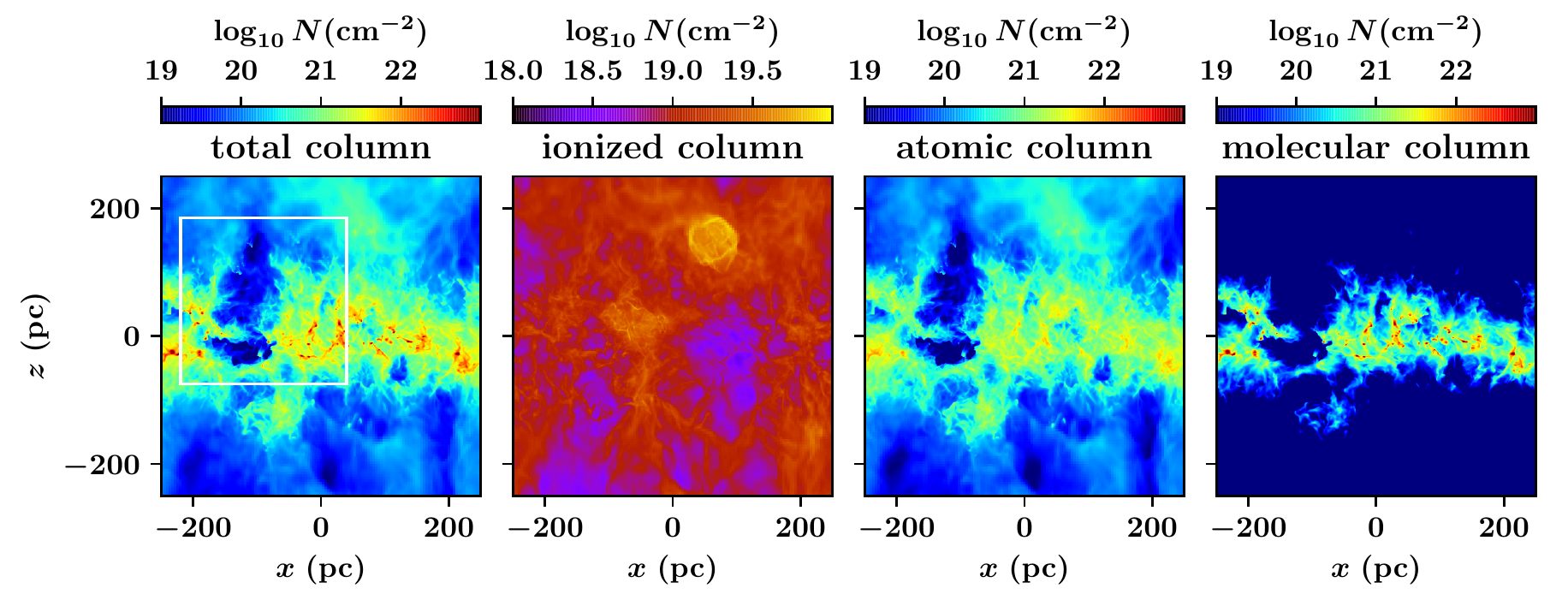}
      \caption{Same as Figure \ref{fig:models} but showing only the column
        density for the three considered gas phases. The white rectangle
        marks the region comparable to the observation. The column density
        range for the ionized gas is much smaller than for the other two
        components. The bright spot in the ionized column coresponds to a
        recently exploded supernova.}\protect\label{fig:models2} 
      \end{center}
   \end{figure*}
The first check against simulations concerns the density distribution in the
region of increased velocity dispersion. Figure \ref{fig:models} shows one of
the frequently appearing feedback features that is comparable to our
observations. From left to right we show the total (ionized, neutral, and
molecular) column density, the modulus of the line-of-sight (los) velocity, the los velocity
dispersion, the thermal sound speed, and the Mach number. The top panels are
the edge-on view of the galactic plane, the bottom ones depict the face-on
view. Shown are the data after a total simulation time of 30 Myr, with the
cluster formation time around $\sim10$ Myr. The white rectangle illustrates
the region that is likely comparable to the observational field. We clearly
note a region of low density, which has been evacuated by clustered stellar
feedback in the disc. This cavity shows higher line-of-sight velocity and corresponding
dispersion, as well as patchy enhanced sound speeds due to higher temperatures,
all displaying values covering the range of our observations. The dispersion
and sound speed increase in a correlated manner such that the Mach number only
shows comparably small variations, quantatively similar to our observations in figures
\ref{fig:sound}a, \ref{fig:sound}c, and \ref{fig:sound}d. Simultaneously, the column density
shows on average a factor of $\sim4$ dex variations between the disk
of the galaxy and the outflow region. Applied to our observations, even
assuming all of the H {\small I} column density of $\sim3\times10^{21}$
cm$^{-2}$ is in front of Mkr 71, which is unlikely, such a substantial density
drop in the outflow region could imply column densities approaching $\sim3$
LyC optical depths, and hence a non-zero LyC escape fraction.

The column density shown in Figure \ref{fig:models} comprises the total
density of gas, including ionized, neutral and molecular. In Figure \ref{fig:models2} we 
demonstrate that the density drop is dominated by the distribution of neutral
H {\small I} - the gas phase of greatest importance for the fate of LyC
photons. We also note that the simulations assume solar metallicity, while the
metallicity of Mrk 71 is substantially subsolar at
$12+\log{\mathrm{O/H}}=7.89\pm0.01$ \citep{Izotov1997,Hunter1999}. However,
for SNe the dynamical impact is not expected to be vastly different at lower
metallicities. The energy injection would be the same, while simultaneously
the cooling would be less efficient, which means overall more of the thermal
energy could be converted into motions and the gas would stay warmer compared
to the solar metallicity case.  

The second hypothesis to test against simulations is related to the orientation of the cone with respect
to the galactic plane of NGC 2366. How likely is the possible outflow cone
pointing away from the plane? The numerical simulations suggest that enhanced
coherent motions are more likely to appear out of the plane. A visual
impression is evident by comparing the top and the bottom panels of Figure
\ref{fig:models}. In both the upper and lower panels the low-density regions
coincide with overall higher los velocities and velocity dispersions. However,
for the lower panels, the correlation is much stronger indicating that the gas
motions are much stronger perpendicular to the plane. A more quantitative
analysis of this phenomenon is shown in \citet[][their figure 6]{Girichidis2018},
over an evolutionary time of 60 Myr and multiple simulations. 
These authors show that for all simulations and over the entire simulation time, the
isotropic motions are noticeably lower than the computed averages, which
proves that the velocities pointing perpendicular to the galactic plane are
dominating over the ones in the plane. Therefore, the interpretation that the observed
cone in Figure \ref{fig:machangle} is oriented out of the plane of the disk
seems reasonable.  

Additionally, the observed ionized outflow is not likely to be powered by an AGN. A
\citet[][BPT]{Baldwin1981} diagram can be used to distinguish between
star-forming galaxies (SFGs) and active galactic nuclei (AGN). We cannot
construct such a diagram due to the lack of [O {\small III}] and \hb\ (i.e.,
the y-axis of a BPT diagram), however, we can still examine the [N {\small
    II}] $\lambda6584$/\ha, and [S {\small II}] $6716+6731$/\ha\ values (i.e.,
the x-axis of a BPT diagram). In a BPT diagram, AGN are typically found at
$\log_{10}$[N {\small II}]/\ha$\gtrsim0.0$ and $\log_{10}$[S {\small
    II}]/\ha$\gtrsim0.0$. To be ambiguous and lie in the ``composite'' SFG/AGN
region, these ratios would have to be $\gtrsim-0.6$. In our PMAS data, all spaxels in the entire
field of view have $\log_{10}$[N {\small II}]/\ha$\lesssim-0.89$ and
$\log_{10}$[S {\small II}]/\ha$\lesssim-0.6$. The ``core'' region has
$\lesssim-1.0$ and $\lesssim-1.6$, respectively, with averages $-1.96$ and
$-1.38$. Such ratios place every spaxel in the extreme end of the BPT diagram,
occupied by low metallicity SFGs. The same is true for line ratios, integrated
over the entire Mrk 71 region, as shown in \citet{Micheva2017b}. The BPT
diagram in \citet{James2016}, constructed from narrowband HST imaging,
similarly shows all pixels to be well to the left of the
theoretical SFG-AGN division line of \citet{Kewley2001}. It is therefore
unlikely that non-thermal radiation from an AGN is significantly contributing to
the ionization of the gas, and is powering the outflow. 

In summary, the kinematics and physical properties of the
observed conical outflow region are much in line with being due to stellar feedback, are consistent with a significant drop in neutral gas density of $\lesssim4$ dex,
and the outflow is likely pointing away from the disk. This is consistent with LyC escape, and
under these conditions LyC photons can directly leak through the ISM, without the
necessity for a ``picket fence'' ISM \citep{Heckman2001,Bergvall2006}, in
which an otherwise optically thick ISM has a covering fraction of less than unity. 

\subsection{The broad kinematic component}

\begin{figure}
     \begin{center}
     \includegraphics[width=\hsize]{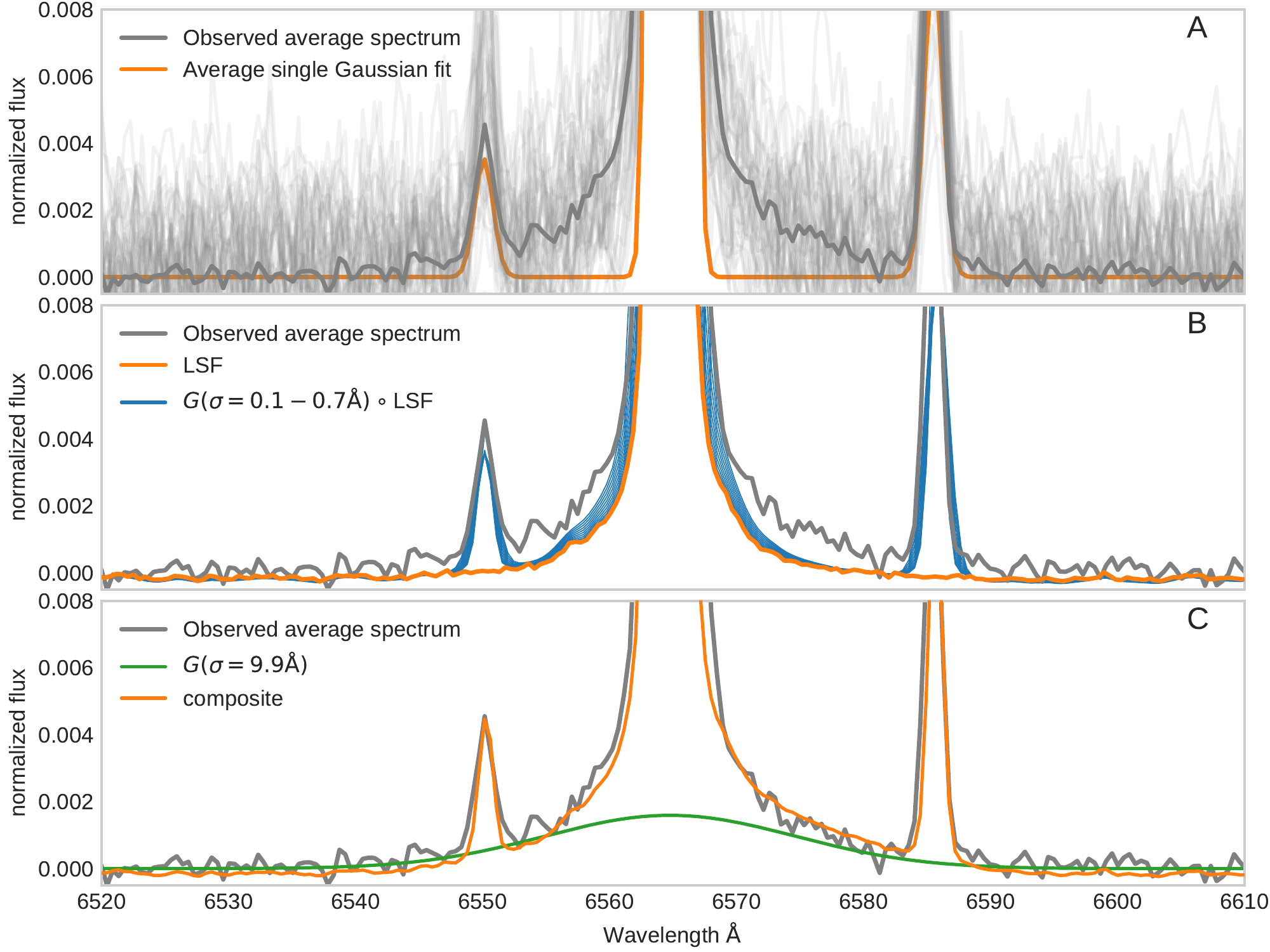}
      \caption{``West'' region average-stacked observed spectrum (thick
        gray line) around \ha, compared to (A) average-stacked single Gaussian models
        (orange line) and individual observed spectra (thin gray lines); (B)
        Empirical LSF (orange line) and LSF-convolved single Gaussian of
        various widths (blue lines); (C) composite model (orange line) of a
        broad Gaussian (green line) added to an LSF-convolved narrow Gaussian
        of width $\sigma=0.7$ \AA. 
        }\protect\label{fig:twocomp_west}
      \end{center}
   \end{figure}

In five GPs, \citet{Amorin2012} find broad H$\alpha$ wings with complex, multi-Gaussian
components, indicating high-velocity gas (FWZI$>1000$ km s$^{-1}$). These
wings are interpreted as a combination of multiple expanding supernovae
remnants and strong stellar winds. Despite the much smaller distance to Mrk
71, the origin of the broad component identified in Section \ref{sec:broad}
is unknown. Our attempt to spatially trace it in individual spectra was
unsuccessful, despite the high S/N in the \ha\ line. The reason for this is
that relative to the peak of the \ha\ line, the broad component seems to be at
the $0.2\%$ level, and we do not have the S/N to trace such a faint component
in most individual spaxels. We therefore average stack all spectra inside of
the ``West'' region, which was identified in Section \ref{sec:broad} as
manifesting the strongest broad component. The resulting average-stacked
spectrum is shown in Figure \ref{fig:twocomp_west} as a solid thick gray line
in all three panels. Here we have normalized all shown spectra to an amplitude
of unity in \ha. In Figure \ref{fig:twocomp_west}a, we examine the possibility
that the observed broadening is a result of the presence of a velocity 
field. Stacking spectra with line peaks offset from $\lambda_\mathrm{rest}$
may result in artificial broadening of the final spectrum (so-called ``beam
smearing''). To test this, we model each \ha\ line with a Gaussian, and then
average stack the modeled profiles, to obtain the solid orange line in Figure
\ref{fig:twocomp_west}. The two [N \small{II}] lines are modeled simulteneously as \ha. Clearly, the resulting profile is much narrower than
the observed \ha\ line (solid thick gray line), and the presence of a velocity
field is not the reason for the broadening. In this same figure we also
overplot the individual spectra (solid thin gray lines), which, while much
noisier, all indicate a broadening wider than the averaged Gaussian model.

   \begin{figure}
     \begin{center}
     \includegraphics[width=\hsize]{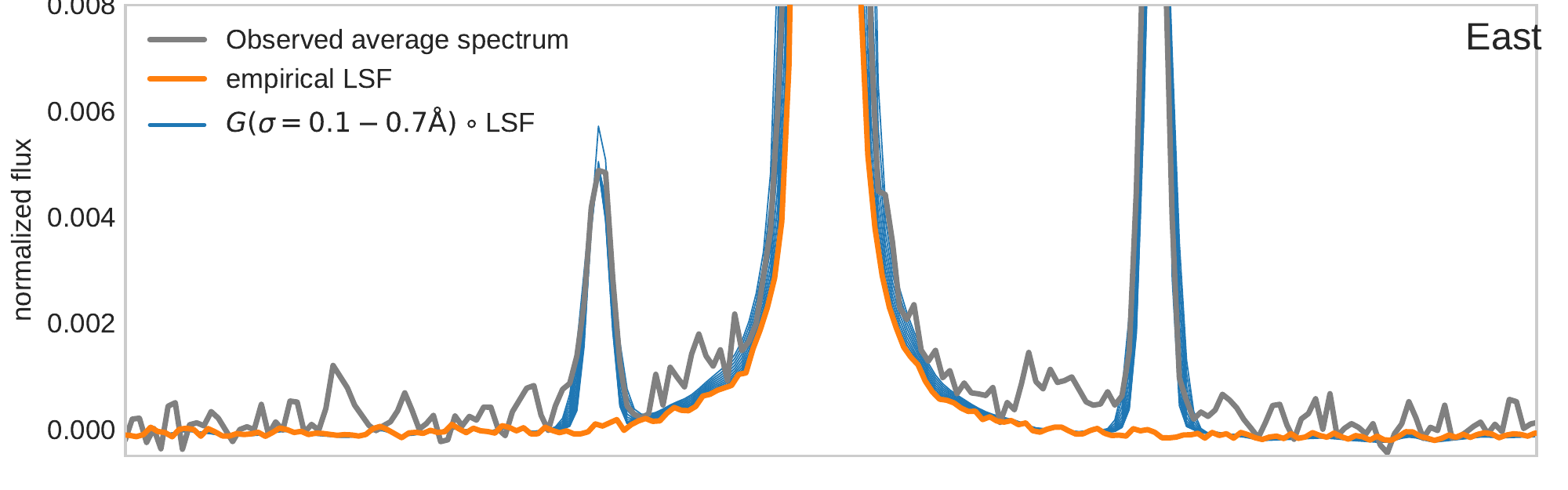}
      \caption{Same as Figure \ref{fig:twocomp_west}b but for the ``East''
        region. No broad component is detected. 
        }\protect\label{fig:twocomp_east}
      \end{center}
   \end{figure}

Next, in Figure \ref{fig:twocomp_west}b we consider the effect of the line
spread function (LSF). We use average-stacked arc lamp lines at $6532.88$ \AA\ and
$6598.95$ \AA\ from the same spatial region (``West''), and obtain an empirical LSF (solid
orange line) at approximately the \ha\ wavelength by average stacking the two arc
lines. We note that the wings of the LSF are much more Lorentzian than
Gaussian. The effect of this LSF on a Gaussian profile can be seen by
convolving Gaussian profiles with widths $\sigma=0.1$ to $0.7$ \AA\ with the
LSF, shown as solid blue lines in Figure \ref{fig:twocomp_west}b. This is done for the \ha\ and the two [N \small{II}] lines simultaneously. None of the
convolved profiles can account for the observed \ha\ wings below an amplitude
of $\lesssim 0.003$, or $0.3\%$ of the \ha\ amplitude.  

Finally, in Figure \ref{fig:twocomp_west}c, we show a composite profile,
consisting of a narrow Gaussian with $\sigma=0.7$ \AA\ for the three lines (\ha\ and two [N \small{II}] lines), convolved with
the LSF, and added to a broad Gaussian of $\sigma=9.9$ \AA. This composite
profile, shown in solid orange, is our best match to the observed
\ha\ profile. The match is not perfect, but it is nevertheless clear that a
broad component does indeed exist, and that \ha\ can be modeled with a
two-component profile. 

In contrast, the ``East'' region, containing knot B, can be modeled with a
single Gaussian, convolved with the LSF, as demonstrated in Figure
\ref{fig:twocomp_east}. The average-stacked observed profile (solid gray line)
is noisier here, and there appears to be some excess emission, localized
around the [N {\small II}] $6584$ line. This is likely a residual from the sky
subtraction. In any case, a single Gaussian component appears to be sufficient
to reproduce the profile of the observed \ha\ line. 

In summary, we can confirm the presence of a broad component in H$\alpha$, 
previously reported by, e.g., \citet{Roy1991}. The averaged broad component in
our data is highly localized to the ``West'' region, and is not broader than
$\mathrm{\sigma\sim10}$ \AA, or a FWHM $\approx24$ \AA $\sim1100$ km/s.
This is narrower than the original detection by \citet{Roy1992},
which had a FWHM of $\lesssim2400$ km s$^{-1}$, and was later re-measured by
\citet{GonzalezDelgado1994} with $\mathrm{FWHM}=1400$ km s$^{-1}$ in the center at a
$3\%$ level of the narrow component, and $\sim2800$ km s$^{-1}$ in the
outskirts at a $30\%$ level. The reason for the discrepancy with our value
($\sim1100$ km/s at a $<0.3\%$ level of the narrow component) is
not certain, however, we can note that \citet{GonzalezDelgado1994} model the
emission lines with Gaussians and do not mention accounting for the effect of
the LSF, which may be non-Gaussian, as it is in PMAS. The insufficient S/N in
the outskirts of our field of view prohibits the spatial 
tracing of this component on a spaxel to spaxel scale. This makes its further
analysis difficult, and we remain ignorant of its origin and nature. We note,
that in contrast to the complex fits with multi-Gaussian components, required
to explain the gas kinematics of GPs in \citet{Amorin2012}, in Mrk 71 it is
not necessary to add more than two Gaussian components to reproduce the
observations. This is expected, since the multiple kinematical components in
the GPs are likely due to multiple star-forming clumps \citep{Amorin2012},
while in Mrk 71 we resolve single star-forming regions.


\subsection{Low- and high-ionization gas move in concert}\protect\label{sec:gas_move}
Finally, we examine if different species, tracing neutral, low- and high-ionization
gas, have the same velocities, using H$\alpha$ as a reference line. If this is
not found to be the case, this could imply local departures from 
thermal equilibrium, in which the rate of collisional excitation does not
equal the recombination rate, or a projection effect of the 2D flattening of a
complex 3D structure. 

\begin{figure}
     \begin{center}
       \includegraphics[width=\hsize]{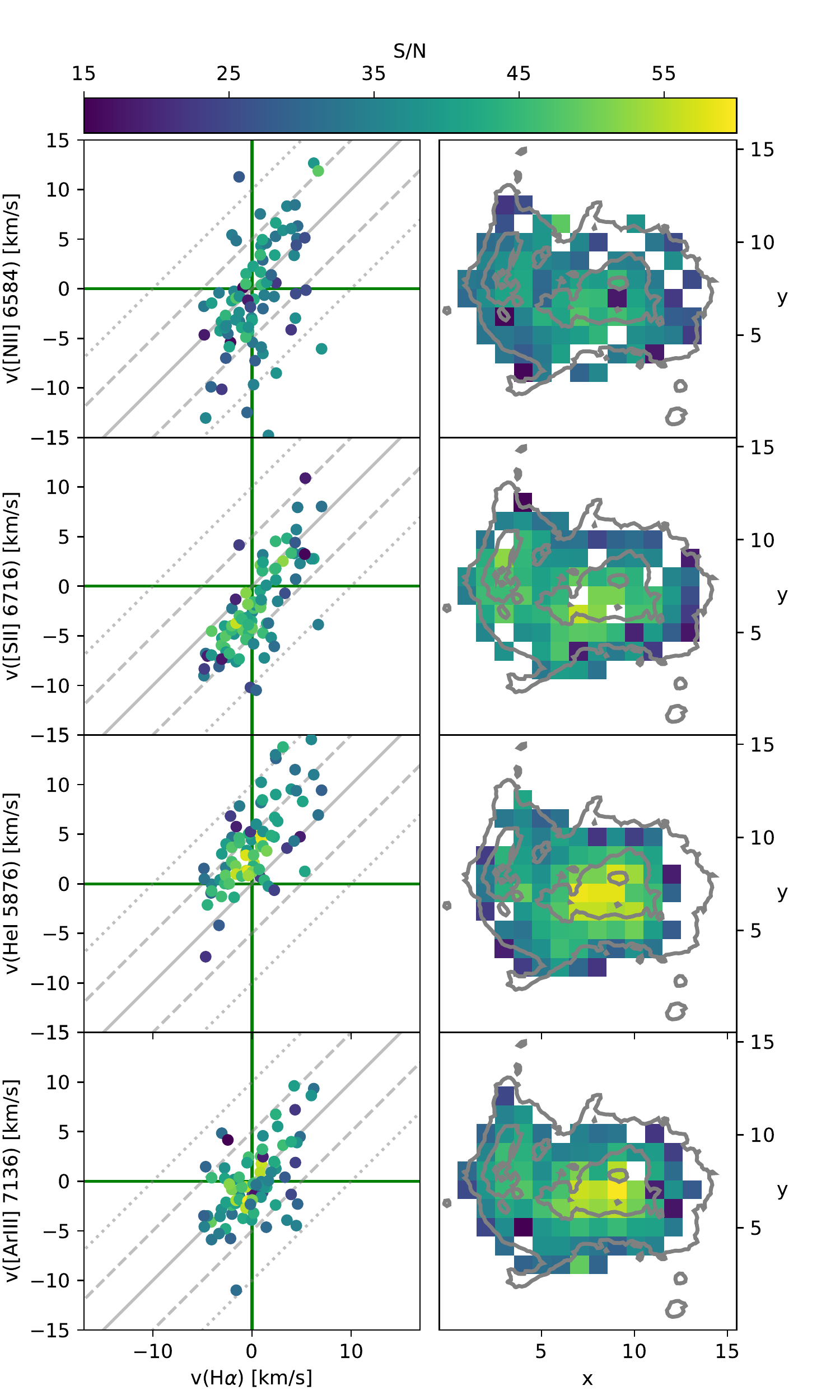}
       \caption{Left: H$\alpha$ radial velocity as a function of the radial
         velocity in lines tracing low-ionization ([N {\small II}], [S {\small
             II}]) and high-ionization gas (He {\small I}, [Ar {\small
             III}]). The points are color-coded by S/N. The $1:1$
         H$\alpha$ velocity line (solid gray), with $\pm5$ (dashed) and
         $\pm10$ km/s (dotted) lines, are plotted for orientation. Only spaxels with
         S/N$\ge15$ and/or within the overplotted contours in the right panel are shown. Right:
         Spatial distribution of the spaxels, color-coded by S/N.}\protect\label{fig:gas_move} 
     \end{center}
   \end{figure}

For every spaxel in the field of view, in Figure \ref{fig:gas_move} we plot
the velocity obtained from the \ha\ line, versus the velocity of the
low-ionization lines [N {\small II}] $6584$ and [S {\small II}] $6716$, and
the high-ionization lines He {\small I} $5876$ and [Ar {\small III}]
$7136$. These lines require ionizing photons with energies $\ge1.1$ and
$\ge0.8$, $\ge1.8$, and $\ge2.0$ Ryd, respectively. Singly-ionized sulfur [S
  {\small II}] therefore traces the kinematics of neutral hydrogen, although
its ionization potential is close to $1$ Ryd, and hence it will also be found
in low-ionization regions together with H$\alpha$. Most data points in Figure
\ref{fig:gas_move} fall within $\pm5$ km/s from the \ha\ velocity. Such small
offsets of $\Delta v\lesssim5$ km/s from the $1:1$ line can be attributed to a
combination of the following systematics. The line spread function (LSF)
varies across the field of view, with the profiles becoming more pronouncedly
non-Gaussian towards the outskirts, which affects the peak position of their
Gaussian models. We attempted to correct for this by modelling these
instrumental systematics from HgNe arc lines in both the wavelength direction
and in the two spatial directions, but found the S/N to be too low in
individual spaxels of the arc line spectra in order to provide a significant
decrease in scatter around the $1:1$ line in Figure \ref{fig:gas_move}. The
uncertainty in the wavelength calibration additionally contributes from
$\pm2.5$ to $\pm5$ km/s uncertainty at different wavelengths, as described in
Section \ref{sec:reduction}. Another consideration is that, other than the
\ha\ line, the S/N of all other lines is significantly lower, making the
determination of the Gaussian peak less accurate. This issue is compounded by
the fact that our direct sky subtraction increased the noise significantly,
making the most difference in spectra with already low S/N. At our minimum
cutoff of S/N$\ge15$, the expected contribution to the uncertainty in the
velocity is $\sim2.5$ km/s, judging by the estimate of \citet[][their figure
  4]{Herenz2016} for the LARS sample with an instrumental setup identical to
ours. Finally, an additional source of scatter is possible stellar continuum
absorption in certain regions of Mrk 71. While knot A is dominated by nebular
continuum and would not be affected, knot B is a naked cluster with a  stellar
continuum clearly detected in, e.g., \citet{Drissen2000} and
\citet{Sokal2016}.  Other spaxels may also be covering regions, where the
stellar continuum is not negligible. In such cases, underlying stellar
absorption lines will influence the position of the peak amplitude of nearby
emission lines.  

Therefore, we conclude that within the effects of instrumental systematics and
possible stellar absorption, gas in different phases shows the same
velocities. In particular, the data is consistent with neutral hydrogen gas
moving together with the ionized gas, and hence the outflow detected in Mrk 71 via
\ha\ and [O {\small III}] $5007$ \citep{Roy1991, GonzalezDelgado1994} is
likely also present in \hi. This is consistent with the simulation predictions
in Figure \ref{fig:models2}.

%

\section{Conclusions}\protect\label{sec:conclusions}

We present $5825$ to $7650$\AA\ PMAS IFU data of the giant H {\small II} region Mrk 71 in NGC 2366,
the closest Green Pea analog at a distance of only $3.4$ Mpc. We obtain
electron density and temperature maps, where for the latter we use a second
order contamination by the [O {\small II}] $\lambda3727$ doublet.

We examine the H$\alpha$
kinematics, and find evidence of a biconical outflow, with elevated velocity dispersion and redshifted (blueshifted)
velocities to the north (south) of its apparent origin at the super star
cluster knot B. An I$\mbox{-}\sigma$ diagnostic verifies both regions as
outflows/expanding shells. 

We verify and locate the previously detected broad velocity component in
H$\alpha$ from stacked spectra, and obtain a maximum full width at half
maximum of $24$\AA, or $1100$ km s$^{-1}$, at a maximum intensity of $\sim0.2\%$
of the narrow component. This is slower and fainter than the previous
detections by \citet{Roy1992} and \citet{GonzalezDelgado1994}. This component
appears to be decoupled from the outflow regions, and its nature remains
unknown, since it cannot be traced by individual spaxels.

We construct spatially resolved sound speed, thermal velocity dispersion,
``true'' velocity dispersion, and Mach number maps, and find evidence for a
gas density drop outside of the core Mrk 71 region. These observational
results are compared to high-resolution SILCC simulations, to demonstrate that
the observed decrease in the gas density may be as large as $\lesssim4$ dex,
even though the associated distribution in Mach numbers shows only minor
differences between outflow and core regions. This implies almost optically
thin H {\small I} column densities, and hence a non-zero LyC escape fraction
in the direction of the outflow. This escape scenario does not require a H
{\small I} covering fraction less than unity in order to enable LyC escape. We
also verify that, within the systematic uncertainties, low- and
high-ionization gas display similar velocities, and therefore likely move
together.  
 
Our results strongly indicate that kinematical feedback is an important
ingredient for LyC leakage in GPs.

\begin{acknowledgements}
      We would like to thank Alaina Henry for refereeing this manuscript. We thank Christer Sandin, Lutz Wisotzki, Norberto Castro,
      Peter Weilbacher, Jakob Walcher, and Tanja Urrutia for enlightening
      discussions during the making of this paper. GM acknowledges support
      from the Leibniz-Wettbewerb, grant SAW-2016-IPHT-2. G\"O acknowledges support from the Swedish Research Council (Vetenskapsr\aa det) and the Swedish Space Board (Rymdstyrelsen). PG acknowledges funding from the European Research Council under ERC-CoG grant CRAGSMAN-646955.
\end{acknowledgements}

\appendix

\section{UV flatfield}
\begin{figure}
     \begin{center}
     \includegraphics[height=7cm]{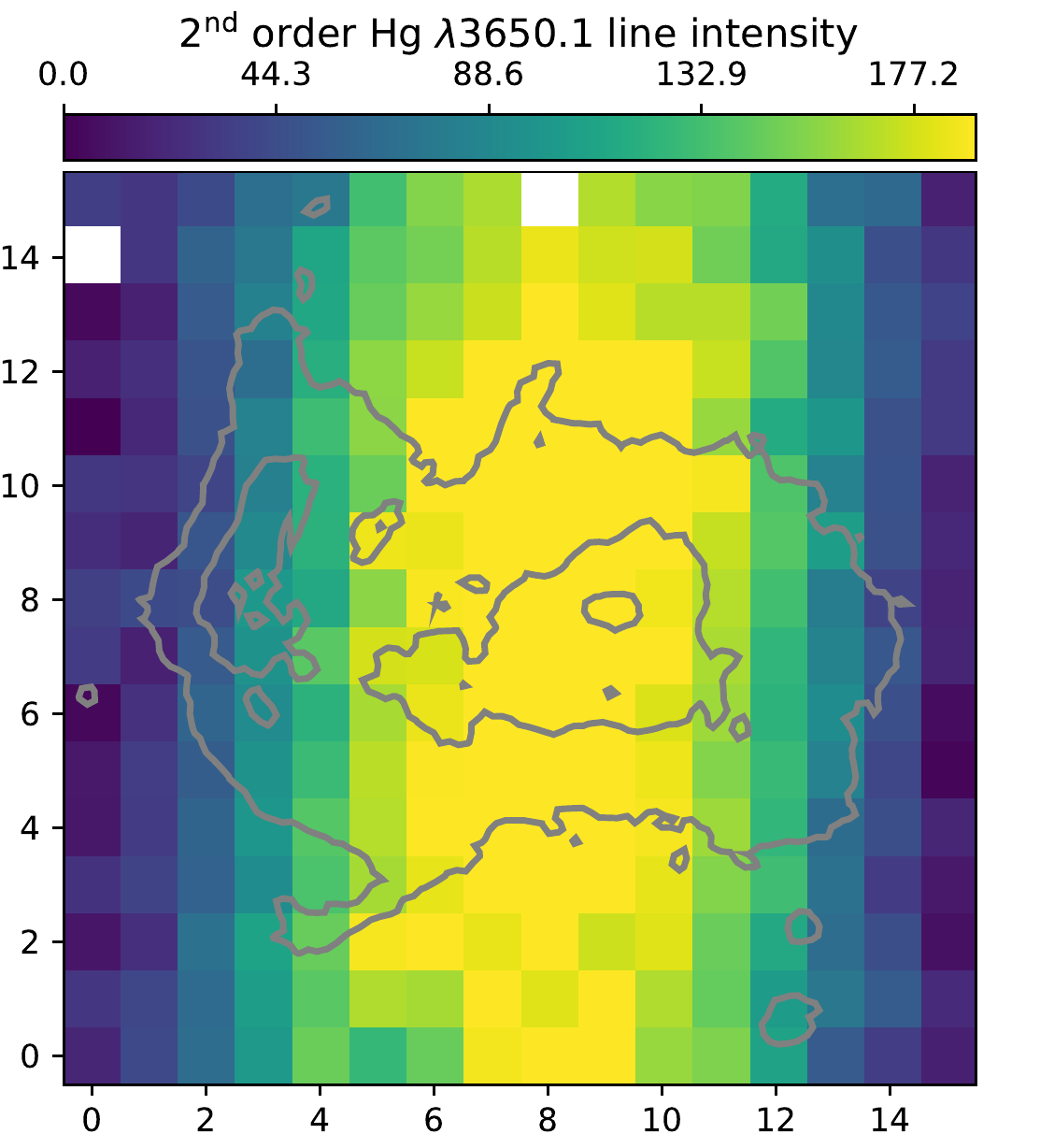}
      \caption{Intensity map of the second order UV Hg line at $\lambda3650$
        \AA, observed at $7300$\AA, and used to obtain a UV
        flatfield. H$\alpha$ contours of Mrk 71 are overplotted for
        orientation.}\protect\label{fig:UV_hg}
      \end{center}
   \end{figure}

\protect\label{sec:UVflat}
In figure \ref{fig:UV_hg} we show the intensity map of a 2nd order spectrum of
the arc lamp line Hg $\lambda3650$ \AA, which is used to obtain a UV flatfield
for the second order [O {\small II}] $\lambda 3727$ \AA\ intensity map.
%

\section{Continuum subtraction in narrowband filters}
\begin{figure*}
     \begin{center}
       \includegraphics[width=\hsize]{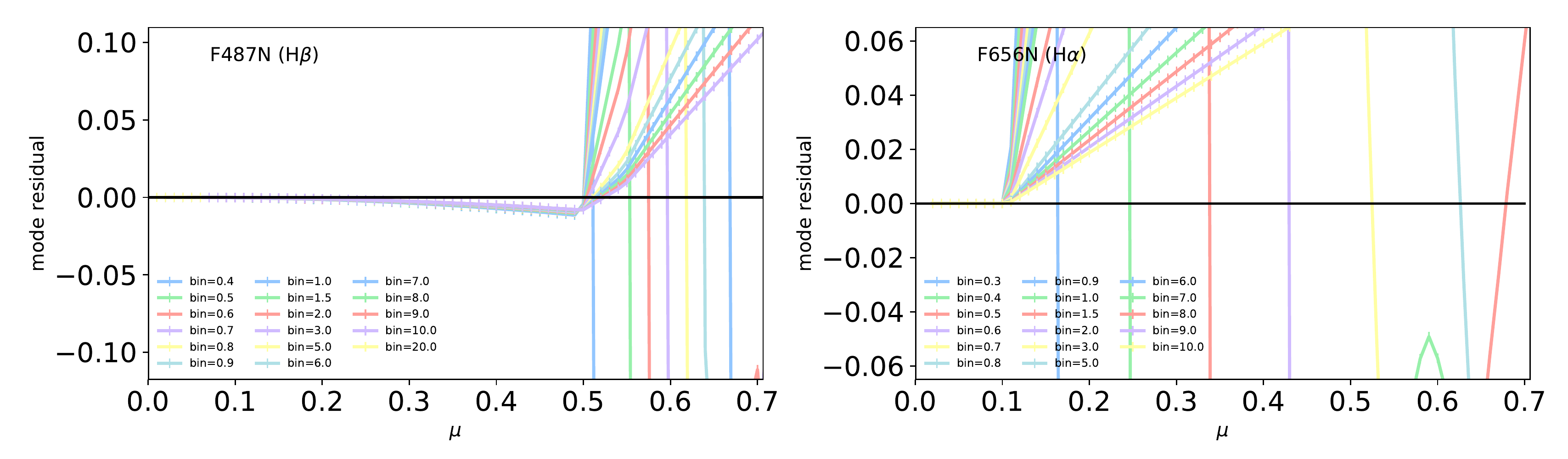}
       \caption{Continuum scaling factor $\mu$ as a function of the mode residual
         for the two HST filters F487N (\hb, left) and F656N (\ha, right). The
         breaks in the two functions indicate a scaling factor of $\mu_S=0.5$
         for \hb\ and $\mu_S=0.1$ for \ha.}\protect\label{fig:mu}
     \end{center}
   \end{figure*}

\begin{figure}
     \begin{center}
       \includegraphics[width=\hsize]{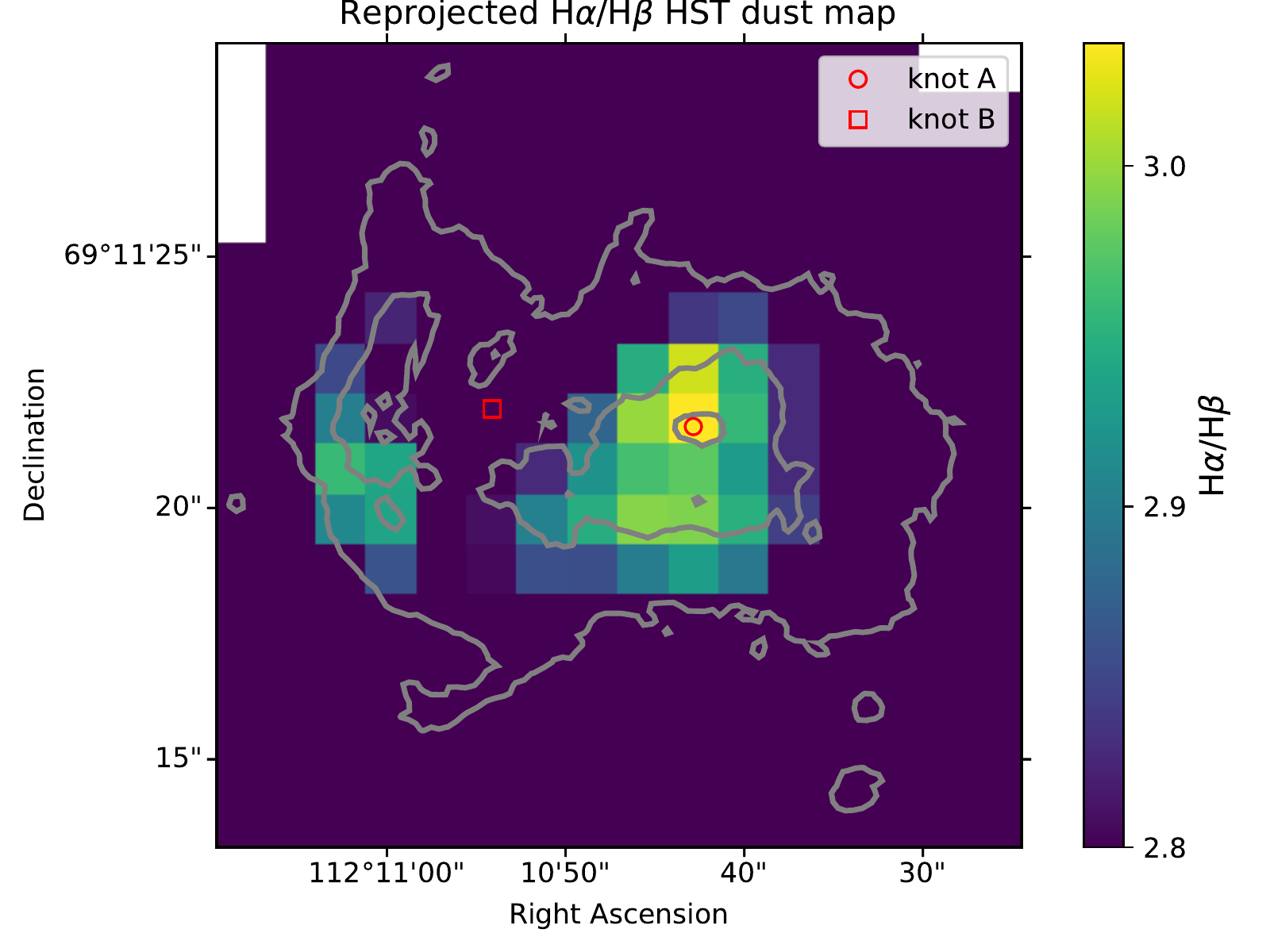}
       \caption{Seeing-convolved, reprojected reddening map from HST
         data. The theoretical $\mathrm{H\alpha/H\beta=2.86}$ at $10^4$ K
         \citep{Brocklehurst1971}, and so unphysically low values imply
         dust-free regions. Additionally, for display purposes spaxels with signal-to-noise $\lesssim2$ are set to the lowest shown value of 2.80. H$\alpha$ contours are overplotted for
         orientation.}\protect\label{fig:dustmap} 
     \end{center}
   \end{figure}

\protect\label{sec:dustmap}
We use the mode method \citep{Keenan2017} to estimate the continuum scaling
factor for the \ha\ and \hb\ HST filters. This method is similar to the
skewness method of \citet{Hong2014}, but uses the mode instead of the
skewness, and performs equally well as other integrated scaling factor
methods, as demonstrated in \citet{Micheva2018}. This is an iterative method,
based on evaluating the mode of the pixel histogram of the
continuum-subtracted image as a function of different scaling factors
$\mu_S$. The mode of the pixel histogram of an image should decrease smoothly as the continuum is subtracted. Once the continuum becomes over-subtracted, the negative pixels will cause a rebinning of the pixel histogram, and a change of the relation between mode and $\mu_S$. This manifests itself as a break in an otherwise smooth function. Since the break indicates the transition from undersubtraction to oversubtraction of the continuum, it marks the location of the optimal scaling factor. The mode of a histogram depends on the bin size used to make that histogram, and therefore one has to explore several bin sizes. Beyond a certain bin size, the behavior of the mode as a function of scaling factor converges to produce a break at the same position for all larger bin sizes. 
For each filter, the pixel histogram is computed over an  area covering
the entire galaxy. Identification of the best scaling factor is done
graphically, as illustrated in figure \ref{fig:mu} for both \ha\ and
\hb\ continuum filters. In these plots of scaling factor $\mu_S$ vs. mode
residual, the break occurs in a single location, and we easily identify
$\mu_S=0.1$ for \ha\ continuum, and $\mu_S=0.5$ for \hb\ continuum. Scaling
factors below and above these values undersubtract and oversubtract the
continuum, respectively. The solution seems quite stable, as indicated by the
large range of bins producing the same break. In figure \ref{fig:dustmap} we
show the resulting \ha/\hb\ map, convolved to the seeing of our PMAS data, and
resampled to the PMAS resolution of $1$ arcsec per spaxel.   

We use the same method to subtract the continuum in the HST F373N from the
archive (PID 13041, PI: B. James), using the F336W filter from the same
run. The corresponding scaling factor is $\mu_S=0.75$. The
continuum-subtracted F373N image is used to calibrate the UV [O {\small II}]
$3726+3729$ second order emission lines in Section \ref{sec:UVOII}. 
%

%
%
\bibliographystyle{aa}
\bibliography{mrk71kinematics}

\label{lastpage}

\end{document}